\documentstyle[12pt]{article}
\newcommand{\nc}{\newcommand}
\nc{\renc}{\renewcommand}

\nc{\etal}{\mbox{\it et al.}}
\nc{\ie}{{\it i.e. }}
\nc{\eg}{{\it e.g. }}

\renc{\thefootnote}{\arabic{footnote}}
\nc{\capt}[1]{{\bf Figure.} {\small\sl #1}}

\nc{\ave}{\bar{E}}
\nc{\avq}{\bar{Q}}
\nc{\avn}{\bar{N}}

\nc{\eqs}[2]{\mbox{Eqs.~(\ref{#1},\,\ref{#2})}}
\nc{\eq}[1]{\mbox{Eq.~(\ref{#1})}}

\nc{\figs}[2]{\mbox{Figs.~(\ref{#1},\,\ref{#2})}}
\nc{\fig}[1]{\mbox{Fig~.(\ref{#1})}}

\nc{\tag}[1]{\label{#1} \marginpar{{\footnotesize #1}}}
\nc{\mtag}[1]{\label{#1} \mbox{\marginpar{{\footnotesize #1}}}}
\renc{\baselinestretch}{1.2}
\newlength{\overeqskip}
\newlength{\undereqskip}
\nc{\be}[1]{\begin{equation} \mbox{$\label{#1}$}}
\nc{\bea}[1]{\begin{eqnarray} \mbox{$\label{#1}$}}
\nc{\Section}[2]{\section{#2}\label{#1}}
\nc{\Bibitem}[1]{\bibitem{#1}}
\nc{\Label}[1]{\label{#1}}

\nc{\eea}{\vspace{\undereqskip}\end{eqnarray}}
\nc{\ee}{\vspace{\undereqskip}\end{equation}}
\nc{\bdm}{\begin{displaymath}}
\nc{\edm}{\end{displaymath}}
\nc{\dpsty}{\displaystyle}
\nc{\bc}{\begin{center}}
\nc{\ec}{\end{center}}
\nc{\ba}{\begin{array}}
\nc{\ea}{\end{array}}
\nc{\bab}{\begin{abstract}}
\nc{\eab}{\end{abstract}}
\nc{\btab}{\begin{tabular}}
\nc{\etab}{\end{tabular}}
\nc{\bit}{\begin{itemize}}
\nc{\eit}{\end{itemize}}
\nc{\ben}{\begin{enumerate}}
\nc{\een}{\end{enumerate}}
\nc{\bfig}{\begin{figure}}
\nc{\efig}{\end{figure}}
\nc{\arreq}{&\!=\!&}
\nc{\arrmi}{&\!-\!&}
\nc{\arrpl}{&\!+\!&}
\nc{\arrap}{&\!\!\!\approx\!\!\!&}
\nc{\non}{\nonumber}
\nc{\align}{\!\!\!\!\!\!\!\!&&}

\def\lsim{\; \raise0.3ex\hbox{$<$\kern-0.75em
      \raise-1.1ex\hbox{$\sim$}}\; }
\def\gsim{\; \raise0.3ex\hbox{$>$\kern-0.75em
      \raise-1.1ex\hbox{$\sim$}}\; }
\nc{\DOT}{\hspace{-0.08in}{\bf .}\hspace{0.1in}}
\nc{\Laada}{\hbox {$\sqcap$ \kern -1em $\sqcup$}}
\nc\loota{{\scriptstyle\sqcap\kern-0.55em\hbox{$\scriptstyle\sqcup$}}}
\nc\Loota{{\sqcap\kern-0.65em\hbox{$\sqcup$}}}
\nc\laada{\Loota}
\nc{\qed}{\hskip 3em \hbox{\BOX} \vskip 2ex}

\nc{\real}{{\rm I \! R}}
\nc{\Z}{{\sf Z \!\!\! Z}}
\nc{\complex}{{\rm C\!\!\! {\sf I}\,\,}}
\def\bigid{\leavevmode\hbox{\small1\kern-3.8pt\normalsize1}}
\def\id{\leavevmode\hbox{\small1\kern-3.3pt\normalsize1}}
\nc{\slask}{\!\!\!/}
\nc{\bis}{{\prime\prime}}
\nc{\pa}{\partial}
\nc{\na}{\nabla}
\nc{\ra}{\rangle}
\nc{\la}{\langle}
\nc{\goto}{\rightarrow}
\nc{\swap}{\leftrightarrow}

\nc{\EE}[1]{ \mbox{$\cdot10^{#1}$} }
\nc{\abs}[1]{\left|#1\right|}
\nc{\at}[2]{\left.#1\right|_{#2}}
\nc{\norm}[1]{\|#1\|}
\nc{\abscut}[2]{\Abs{#1}_{\scriptscriptstyle#2}}
\nc{\vek}[1]{{\rm\bf #1}}
\nc{\integral}[2]{\int\limits_{#1}^{#2}}
\nc{\inv}[1]{\frac{1}{#1}}
\nc{\dd}[2]{{{\partial #1}\over{\partial #2}}}
\nc{\ddd}[2]{{{{\partial}^2 #1}\over{\partial {#2}^2}}}
\nc{\dddd}[3]{{{{\partial}^2 #1}\over
    {\partial #2 \partial #3}}}
\nc{\dder}[2]{{{d #1}\over{d #2}}}
\nc{\ddder}[2]{{{d^2 #1}\over{d {#2}^2}}}
\nc{\dddder}[3]{{d^2 #1}\over
    {d #2 d #3}}
\nc{\dx}[1]{d\,^{#1}x}
\nc{\dy}[1]{d\,^{#1}y}
\nc{\dz}[1]{d\,^{#1}z}
\nc{\dl}[1]{\frac{d\,^{#1}l}{(2\pi)^{#1}}}
\nc{\dk}[1]{\frac{d\,^{#1}k}{(2\pi)^{#1}}}
\nc{\dq}[1]{\frac{d\,^{#1}q}{(2\pi)^{#1}}}

\nc{\cc}{\mbox{$c.c.$ }}
\nc{\hc}{\mbox{$h.c.$ }}
\nc{\cf}{cf.\ }
\nc{\erfc}{{\rm erfc}}
\nc{\Tr}{{\rm Tr\,}}
\nc{\tr}{{\rm tr\,}}
\nc{\pol}{{\rm pol}}
\nc{\sign}{{\rm sign}}
\nc{\bfT}{{\bf T }}
\def\GeV{{\rm\ GeV}}

\nc{\cA}{{\cal A}}
\nc{\cB}{{\cal B}}
\nc{\cD}{{\cal D}}
\nc{\cE}{{\cal E}}
\nc{\cG}{{\cal G}}
\nc{\cH}{{\cal H}}
\nc{\cL}{{\cal L}}
\nc{\cO}{{\cal O}}
\nc{\cT}{{\cal T}}
\nc{\cN}{{\cal N}}
\nc{\rvac}[1]{|{\cal O}#1\rangle}
\nc{\lvac}[1]{\langle{\cal O}#1|}
\nc{\rvacb}[1]{|{\cal O}_\beta #1\rangle}
\nc{\lvacb}[1]{\langle{\cal O}_\beta #1 |}
\nc{\bb}{\bar{\beta}}
\nc{\bt}{\tilde{\beta}}
\nc{\ctH}{\tilde{\cal H}}
\nc{\chH}{\hat{\cal H}}
\nc{\1}{\aa}
\nc{\2}{\"{a}}
\nc{\3}{\"{o}}
\nc{\4}{\AA}
\nc{\5}{\"{A}}
\nc{\6}{\"{O}}
\nc{\al}{\alpha}
\nc{\g}{\gamma}
\nc{\Del}{\Delta}
\nc{\e}{\epsilon}
\nc{\eps}{\epsilon}
\nc{\lam}{\lambda}
\nc{\Om}{\Omega}
\nc{\ve}{\varepsilon}
\nc{\mn}{{\mu\nu}}
\nc{\k}{\kappa}
\nc{\vp}{\varphi}

\nc{\advp}[3]{{\it  Adv.\ in\ Phys.\ }{{\bf #1} {(#2)} {#3}}}
\nc{\annp}[3]{{\it  Ann.\ Phys.\ (N.Y.)\ }{{\bf #1} {(#2)} {#3}}}
\nc{\apl}[3]{{\it  Appl. Phys. Lett. }{{\bf #1} {(#2)} {#3}}}
\nc{\apj}[3]{{\it  Ap.\ J.\ }{{\bf #1} {(#2)} {#3}}}
\nc{\apjl}[3]{{\it  Ap.\ J.\ Lett.\ }{{\bf #1} {(#2)} {#3}}}
\nc{\app}[3]{{\it Astropart.\ Phys.\ }{{\bf #1} {(#2)} {#3}}}
\nc{\cmp}[3]{{\it  Comm.\ Math.\ Phys.\ }{{ \bf #1} {(#2)} {#3}}}
\nc{\cqg}[3]{{\it  Class.\ Quant.\ Grav.\ }{{\bf #1} {(#2)} {#3}}}
\nc{\epl}[3]{{\it  Europhys.\ Lett.\ }{{\bf #1} {(#2)} {#3}}}
\nc{\ijmp}[3]{{\it Int.\ J.\ Mod.\ Phys.\ }{{\bf #1} {(#2)} {#3}}}
\nc{\ijtp}[3]{{\it Int.\ J.\ Theor.\ Phys.\ }{{\bf #1} {(#2)} {#3}}}
\nc{\jmp}[3]{{\it  J.\ Math.\ Phys.\ }{{ \bf #1} {(#2)} {#3}}}
\nc{\jpa}[3]{{\it  J.\ Phys.\ A\ }{{\bf #1} {(#2)} {#3}}}
\nc{\jpc}[3]{{\it  J.\ Phys.\ C\ }{{\bf #1} {(#2)} {#3}}}
\nc{\jap}[3]{{\it J.\ Appl.\ Phys.\ }{{\bf #1} {(#2)} {#3}}}
\nc{\jpsj}[3]{{\it J.\ Phys.\ Soc.\ Japan\ }{{\bf #1} {(#2)} {#3}}}
\nc{\lmp}[3]{{\it Lett.\ Math.\ Phys.\ }{{\bf #1} {(#2)} {#3}}}
\nc{\mpl}[3]{{\it  Mod.\ Phys.\ Lett.\ }{{\bf #1} {(#2)} {#3}}}
\nc{\ncim}[3]{{\it  Nuov.\ Cim.\ }{{\bf #1} {(#2)} {#3}}}
\nc{\np}[3]{{\it  Nucl.\ Phys.\ }{{\bf #1} {(#2)} {#3}}}
\nc{\pr}[3]{{\it Phys.\ Rev.\ }{{\bf #1} {(#2)} {#3}}}
\nc{\pra}[3]{{\it  Phys.\ Rev.\ A\ }{{\bf #1} {(#2)} {#3}}}
\nc{\prb}[3]{{\it  Phys.\ Rev.\ B\ }{{{\bf #1} {(#2)} {#3}}}}
\nc{\prc}[3]{{\it  Phys.\ Rev.\ C\ }{{\bf #1} {(#2)} {#3}}}
\nc{\prd}[3]{{\it  Phys.\ Rev.\ D\ }{{\bf #1} {(#2)} {#3}}}
\nc{\prl}[3]{{\it Phys\ Rev.\ Lett.\ }{{\bf #1} {(#2)} {#3}}}
\nc{\pl}[3]{{\it  Phys.\ Lett.\ }{{\bf #1} {(#2)} {#3}}}
\nc{\prep}[3]{{\it Phys\. Rep.\ }{{\bf #1} {(#2)} {#3}}}
\nc{\prsl}[3]{{\it Proc.\ R.\ Soc.\ London\ }{{\bf #1} {(#2)} {#3}}}
\nc{\ptp}[3]{{\it  Prog.\ Theor.\ Phys.\ }{{\bf #1} {(#2)} {#3}}}
\nc{\ptps}[3]{{\it  Prog\ Theor.\ Phys.\ suppl.\ }{{\bf #1} {(#2)} {#3}}}
\nc{\physa}[3]{{\it  Physica\ A\ }{{\bf #1} {(#2)} {#3}}}
\nc{\physb}[3]{{\it  Physica\ B\ }{{\bf #1} {(#2)} {#3}}}
\nc{\phys}[3]{{\it Physica\ }{{\bf #1} {(#2)} {#3}}}
\nc{\rmp}[3]{{\it  Rev.\ Mod.\ Phys.\ }{{\bf #1} {(#2)} {#3}}}
\nc{\rpp}[3]{{\it Rep.\ Prog.\ Phys.\ }{{\bf #1} {(#2)} {#3}}}
\nc{\sjnp}[3]{{\it Sov.\ J.\ Nucl.\ Phys.\ }{{\bf #1} {(#2)} {#3}}}
\nc{\spjetp}[3]{{\it Sov.\ Phys.\ JETP\ }{{\bf #1} {(#2)} {#3}}}
\nc{\yf}[3]{{\it Yad.\ Fiz.\ }{{\bf #1} {(#2)} {#3}}}
\nc{\zetp}[3]{{\it Zh.\ Eksp.\ Teor.\ Fiz.\  }{{\bf #1}  {(#2)} {#3}}}
\nc{\zp}[3]{{\it Z.\ Phys.\ }{{\bf #1} {(#2)} {#3}}}
\nc{\ibid}[3]{{\sl ibid.\ }{{\bf #1} {#2} {#3}}}
\nc{\rf}[1]{(\ref{#1})}
\nc{\nn}{\nonumber \\*}
\nc{\bfB}{\bf{B}}
\nc{\bfv}{\bf{v}}
\nc{\bfx}{\bf{x}}
\nc{\bfy}{\bf{y}}
\nc{\vx}{\vec{x}}
\nc{\vy}{\vec{y}}
\nc{\oB}{\overline{B}}
\nc{\oI}{\overline{I}}
\nc{\oR}{\overline{R}}
\nc{\rar}{\rightarrow}
\nc{\ti}{\times}
\nc{\slsh}{\hskip-5pt/}
\nc{\sm}{Standard~Model~}
\nc{\MP}{M_{\rm Pl}}
\nc{\tp}{t_{\rm Pl}}

\nc{\pmin}{p_{\rm min}}
\nc{\pmax}{p_{\rm max}}
\nc{\fo}{f_0}
\nc{\foi}{f_{0,i}\,}
\nc{\fop}{f_0^P}
\nc{\fou}{f_0^U}

\nc{\eff}{{\rm eff}}
\nc{\MT}{M_{\rm T}}
\nc{\ML}{M_{\rm L}}
\nc{\kk}{\vek{k}}
\nc{\pp}{{\rm p}}
\nc{\pt}{\partial_t}
\nc{\half}{{1\over 2}}
\nc{\w}{\omega}
\nc{\uhat}{\hat{U}_\w}

\begin{document}
{\title{\vskip-2truecm{\hfill {{\small HIP-2000-59/TH\\
  \hfill {\small TURKU-FL-P36-00}  \hfill \\
    }}\vskip 1truecm}
{\LARGE Numerical simulations of fragmentation of the Affleck-Dine
condensate}}
\vspace{-.2cm}
{\author{
{\sc Kari Enqvist$^{1}$}
\\
{\sl\small Physics Department, University of Helsinki,   and Helsinki Institute of Physics}
\\
{\sl\small P.O. Box 9, FIN-00014 University of Helsinki, Finland}
\\
{\sc Asko Jokinen$^{2}$}
\\
{\sl\small Physics Department, University of Helsinki}
\\
{\sl\small P.O. Box 9, FIN-00014 University of Helsinki, Finland}
\\
{\sc Tuomas Multam\"aki$^{3}$ and Iiro Vilja$^{4}$ }
\\
{\sl\small Department of Physics, University of Turku, FIN-20014 Turku, Finland}\\
}
\date{November 9, 2000}
\maketitle
\begin{abstract}
\noindent
We present numerical simulations of fragmentation of the Affleck-Dine
condensate in two spatial dimensions. We argue analytically that
the final state should consist of both Q-balls
and anti-Q-balls in a state of maximum entropy, with most of the
balls small and relativistic. Such a behaviour is found in simulations
on a 100$\times$100 lattice with cosmologically realistic parameter
values. During fragmentation process, we observe filament-like texture
in the spatial distribution of charge. The total charge in Q-balls is 
found to be almost equal to the charge in anti-Q-balls and typically
orders of magnitude larger than charge asymmetry. Analytical
considerations
indicate that, apart from geometrical factors, the results of
the simulated two dimensional case should apply also to the fully realistic
three dimensional case.
\end{abstract}
\vfil
\footnoterule
{\small $^1$kari.enqvist@helsinki.fi;  $^2$asko.jokinen@helsinki.fi;}
{\small $^3$tuomas@maxwell.tfy.utu.fi;}\\
{\small   $^4$vilja@newton.tfy.utu.fi}

\thispagestyle{empty}
\newpage
\setcounter{page}{1}

\section{Introduction}
The Minimal Supersymmetric Standard Model (MSSM) has several flat directions
where the scalar potential is nearly identically zero \cite{dinert}. 
During inflation, squark and slepton fields will fluctuate freely
along the flat directions, forming Affleck-Dine (AD) -condensates  \cite{ad}.
The  state of lowest energy, however, is not the AD-condensate 
but a non-topological soliton, the Q-ball \cite{cole2,ks1}, 
which carries a non-zero baryonic (B-ball) or leptonic (L-ball) charge. 
The instability is induced by the spatial perturbations \cite{ks,bbb1}
which are naturally present in the condensate because of quantum
fluctuations during inflation.
The fragmentation process and
the properties of the resulting Q-balls depend on SUSY breaking:
if gauge mediated, Q-balls will be large and completely stable \cite{ks,ksd}
and form along every flat direction\footnote{For another variant of
gauge mediated Q-balls, see \cite{kasuya3}.}
and could be detectable even today \cite{kkst},
whereas if SUSY breaking is gravity mediated, Q-balls will be 
unstable  \cite{bbb1,bbb2} but long-lived enough to have a host of interesting
cosmological consequences \cite{bbb2,bbbdm}.
In the gravity mediated case the
formation of Q-balls is  also a generic feature in all but few
flat directions \cite{ejm}. 

Fragmentation of the AD-condensate involves highly non-linear
dynamics and is therefore quite complicated \cite{adfrag,mv1}. 
The condensate lumps
must somehow loose the extra energy when
settling down to a Q-ball configuration. This they may do by
radiating out quanta of AD-scalars or smaller condensate lumps,
but the lumps  may also experience frequent collisions.
Since for a fixed charge the energy of a Q-ball is also fixed, 
fragmentation dynamics and the final distribution of Q-balls
will obviously depend on the initial
energy and charge density of the condensate. In principle,
this is a free parameter, although the observed baryon asymmetry
can be argued to indicate a natural order of magnitude for the ratio.

In Ref. \cite{adfrag} the relaxation of a single spherical
condensate lump was followed numerically, but sphericality
is likely to be a far too constraining assumption. Moreover, the condensate
fragments are not isolated but interact with each other.
The general features of 
Q-ball collisions have been studied in Refs. \cite{collisions,mv2,mv3};
the cross section in the gravity mediated
case has been shown to depend on the relative phases of
the colliding Q-balls \cite{mv2}.
Q-ball formation has been seen in 3-d lattice studies both in
gauge mediated \cite{kasuya1} and gravity mediated \cite{kasuya2} cases
but only for a single energy-to-charge ratio of the condensate in each case.
As we will show in this paper, natural choices for the initial 
energy-to-charge ratio necessarily lead to a copious formation of not only
Q-balls but also of anti-Q-balls, which carry a net negative charge.
Such a conclusion can be reached already by relatively simple analytical
arguments which assume that the condensate lump reaction rates are
fast enough to drive them into a state of maximum entropy. From the
resulting equilibrium distributions it then follows that the number
of Q-balls is almost equal to the number of anti-Q-balls so that
the total positive (negative) charge in Q-balls is typically much
larger than the actual charge asymmetry. 
Indeed, as we shall discuss in this paper, detailed numerical simulations
show this behaviour.

The paper is organized as follows.
In Section 2 we derive the Q-ball and anti-Q-ball number density
distributions by assuming that the condensate lumps actually
equilibrate; this is shown to be a self-consistent assumption.
We also show that as far as the distributions are concerned,
there is no qualitative difference between
two and three spatial dimensions.
In Section 3 we present the results of numerical simulations of
the condensate fragmentation in 2+1 dimensions. We follow the
time evolution of the condensate field, show it breaking up
into Q-balls and anti-Q-balls, and extract the number density
distributions from the numerical data. In Section 4 we present
our conclusions.

\section{Distributions of Q-balls and anti-Q-balls}
\subsection{AD-condensate fragmentation}\label{frag}

In the cosmological context the formation of Q-balls via 
fragmentation of the AD-condensate has to be characterized by taking 
into account the expansion of the universe. The random 
perturbations on the homogeneous AD-condensate field generated during
de Sitter -phase of the inflation first 
grow linearly as long as they remain small compared to the AD-field.
When these linear modes are large enough, they become non-linear mode by mode 
and their dynamics becomes much more complicated. During this non-linear
era the actual Q-balls are formed as the system finds the energetically
most favourable configuration ending up with a particular system of Q-balls and, in general,
radiation.
 
To be able to quantify the dynamics of the Q-balls one has to 
specify the potential. It consists of various terms appearing 
from the effective field theory of the particular case. 
To lowest order, a generic
general potential reads
 \be{epot}
V_H(\Phi ) = V(\Phi ) - g H^2 |\Phi |^2,
\ee
where the second  term, depending on the Hubble rate 
$H=\dot a/a$, where $a(t)$ is the scale factor, originates from higher-order operators
in the K\" ahler potential (with typically $g \sim 1$), while
$V$ includes all terms of the flat direction.

Of particular interest are flat-direction potentials of the form \cite{bbb1} 
\be{potEMcD} V(\Phi) = m^2|\Phi |^2(1 + K \log({|\Phi |^2\over 
M^2}))+{\lambda^2\over M^2_{Pl}}|\Phi|^6,
 \ee
arising in the gravity mediated SUSY breaking scenario with a d=4
flat direction, and \cite{ks} 
\be{potKS} V(\Phi ) = m^4 \ln\left (1+ {|\Phi 
|^2\over m^2}\right ) +{\lambda\over M^2_{Pl}} |\phi |^6 \ee 
in 
the gauge mediated SUSY breaking scenario. 
The mass scale $m$ is given by the SUSY breaking scale; typically $m\sim {\cal O}(100)\GeV$.
We have here 
omitted all terms which violate the quantum number $Q$. These are 
surely needed for generating the net charge of the 
AD-condensate, but  the $Q$-violating terms are 
negligible at the time when the condensate finally begins to fragment. 
Therefore we may simply assume as an initial condition that at the onset of fragmentation
the AD-condensate has some non-zero 
initial charge $Q$.

The initial state of the AD-condensate is
determined by two parameters, the energy and charge density of the
condensate.
For a fixed charge $Q$ the critical parameter determining the
Q-ball distribution is the ratio of energy density $\rho$ to charge 
density $q$, 
\be{x} x(\Phi ) = {\rho(\Phi)\over m\,q(\Phi)}. 
\ee 
One should note that according to our assumptions the charge density is a
conserved quantity whereas the energy density is not. However, a
simple calculation reveals that in a matter dominated universe 
the energy pumped into the scalar field by gravitation is very 
small compared to the initial energy of the condensate,
so that in practice also $\rho$ is conserved. Hence, the 
ratio $x$ is a good parameter to describe the system. It is clear 
that the larger $x$ is the more energy should be stored as
radiation, kinetic energy of Q-balls, anti-Q-balls or some 
combination of these.

To get an impression of the general features of the Q-ball formation process,
it is useful to study instabilities in both gauge and
gravity mediated scenarios. By 
writing the AD-condensate in terms of the modulus and the phase, 

\be{ad}
\Phi=\phi\, e^{i\theta},
\ee 
one can formalize the requirements for 
the occurrence of the instabilities $\delta\Phi$ in the model. 
When the Q-balls begin to form, the instability band of 
fluctuations in the gauge mediated case is essentially given by \cite{kasuya1}
\be{ins1}
{k\over a(t)} \lsim  {m^2\over \phi },
\ee 
where it is 
assumed that the field $\phi$ is large. This implies, that the 
size of the instabilities
is bounded below by 
\be{bound1} \lambda \sim \frac 1k \gsim {|\Phi |\over a(t) m^2}. 
\ee 

In a cosmological context where the field is large this implies 
that the Q-balls will be very large. Indeed, this can be seen in 
the study by Kasuya and Kawasaki \cite{kasuya1},
where they establish the formation of Q-balls using 3D lattice 
simulations. In their simulation one can see a single large 
Q-ball containing virtually all of the initial charge set into 
the system: this Q-ball most likely forms from the perturbation first 
entering the lattice, i.e. the first unstable mode which fits 
inside the lattice. However, the most amplified mode is even 
larger. These considerations make the gauge mediated scenario 
ineffective to study on a lattice: the lattice size should be huge 
in order to contain the relevant modes visible already at the very 
beginning.

The situation is, however, different in the gravity mediated 
case. The instability band is characterized by 
\be{ins2} 
{k^2\over a^2} \lsim \dot\theta^2- V'', \ee 
which has to be positive in order to have growing modes at all. Moreover, the 
maximally growing mode can be identified as $(k_{max}/a)^2=\frac 
12 (\dot\theta^2- V'')$ and its growth is characterized by 
$|\delta\Phi| \sim e^{\alpha(t)}$, where 
\be{alpha} 
\dot\alpha (t) = {\dot\theta^2- V'' \over 4\dot\theta }. 
\ee 
If $\dot\theta\simeq m$ the wave number of the maximally growing 
mode is $k_{max}=\sqrt{\frac 32 |K|}\, a(t)\,m$ and $\dot\alpha =\frac 38 
\,|K|\,m$. This would mean that the criteria for the end of linear 
growth, $\left | {\delta\Phi\over\Phi }\right |\sim 1$, implies $ 
\alpha (t)\sim\dot\alpha t\sim\ln\left | {\Phi 
(t=0)\over\delta\Phi (t=0)}\right |$ which, in the light of the 
simulations, is typically too small, as will become evident later. 
Therefore it is important to recognize that for the $x>1$ cases 
$\dot\theta (t=0) \ll m$, and hence no instabilities occur 
because $V''$ is  typically positive. Due to the dynamics 
$\dot\theta$ begins to grow and unstable modes appear when 
$\dot\theta^2- V''$ changes its sign to positive. At that 
time, however, the rate of perturbation growth is very slow 
because $\dot\alpha$ is very small ($\dot\theta\sim m$). So the growth of the
perturbations is very slow at first, but later the rate of growth
increases and becomes much faster when the average $\dot\theta^2$ is 
large enough. This lengthens remarkably the time needed to 
enter the non-linear era of perturbation growth and pushes the
actual formation of Q-balls later. 
 
Note that when the growth of linear modes is finally speeded up, 
the most amplified modes are
 remarkably smaller than in the gauge mediated
scenario. This naturally implies that the size of the largest 
Q-ball is much larger in the gauge mediated than in the gravity 
mediated scenario.
 
In both gauge and gravity mediated cases the unstable 
perturbations $\delta\Phi$ remain linear as long as they are much 
smaller than the background AD-condensate field $\Phi$: 
\be{linehto} 
\left | {\delta \Phi\over\Phi}\right | \ll 1. 
\ee 
On 
the other hand the charge density perturbations $\delta q = 
q(\Phi + \delta\Phi) - q(\Phi)$ are always smaller or equal than the field 
perturbations, i.e. 
\be{linehto2}
 \left | {\delta q\over q}\right 
| \leq \left | {\delta\Phi\over\Phi}\right |. 
\ee 
This implies 
definitely that $ \left | {\delta q\over q}\right | \sim 1$ 
indicates the end of the linear growth era of the perturbations.

\subsection{Equilibrium ensembles}

After fragmentation, the AD lumps are expected to interact vigorously. Gradually the field fragments
will settle to the state of lowest energy by emitting and exchanging smaller fragments.
If the interaction is fast enough compared with the expansion rate of the universe, a natural expectation is
that the final state 
should consist of an equilibrium distribution of Q-balls and anti-Q-balls in a state of maximum
entropy. We shall now work out the consequences of such an assumption in the case of
gravity mediated Q-balls, for which the mass $M_Q$ is given approximately by
\be{qmassa}
M_Q\approx m\,Q~.
\ee

Q-ball (anti-Q-ball) distributions $N_+(Q,p)$ ($N_-(Q,p)$) are subject to the following constrains:
\bea{eqehto}
E_{tot} = E_++E_- ~, & E_{\pm} = \int dQ~dp~E(Q,p)~N_{\pm}(Q,p)~\nn
Q_{tot} = Q_+-Q_- ~, & Q_{\pm} = \int dQ~dp~Q~N_{\pm}(Q,p)~,
\eea
where $E(Q,p)\approx\sqrt{p^2+m^2Q^2}$ is the energy of a single Q-ball, $E_+$ $(E_-)$ and $Q_+$ $(Q_-)$ are 
the energy and charge of Q-balls (anti-Q-balls), and $E_{tot}$ and $Q_{tot}$
are respectively the total energy and charge of Q-balls and anti-Q-balls,
which are equal to the corresponding values of the AD-condensate in the
beginning unless significant amounts of energy and/or charge are transformed
into radiation. It then follows from \eq{eqehto} that
\be{epayht}
E_{tot}\geq m\,(Q_++Q_-) \geq m\,Q_{tot}.
\ee
 This condition is independent of the
precise form of the Q-ball distributions. It is easy to see that the
energy-to-charge  ratio (see \eq{x}) $x\geq1$. If $x=1$, the inequalities 
in \eq{epayht} simplify to equalities and this gives directly the Q-ball
distributions
\be{distrib1} N_+(Q,p) = N_+(Q)~\delta(p)~,\qquad N_-(Q,p) =0~,
\ee
 which means that there are no anti-Q-balls and that
the Q-balls for at rest with a distribution $N_+(Q)$. However, this is not a realistic case. 
Indeed, if all the baryon asymmetry resides in Q- and anti-Q-balls,
then at times earlier than $10^{-6}s$, 
$Q_{tot}/Q_+\sim \Delta B\sim 10^{-8}$. (Since $B-L$ is conserved in
the MSSM, this holds also for the purely leptonic flat directions.)
{}From \eq{epayht} it then would follow that 
\be{xey}
x \simeq 10^8~.
\ee
Even if all of the baryon asymmetry were not carried by Q-balls (or
rather, in this case, by B-balls), it is likely that some of it is.
As a consequence, a natural expectation is 
$x\gg 1$. 

Hence the number of Q-balls, $N_+$, and the number of anti-Q-balls, 
$N_-$, are approximately equal so that the total number of Q-balls 
is $N_{tot}\approx 2N_+$. Then from \eq{eqehto} it follows that
\bea{number1}
E_{\pm} &\approx& E(\bar{Q},\bar{p})~\int dQ~dp~N_{\pm}(Q,p)~\approx~
 E(\bar{Q},\bar{p})~N_+~, \nonumber \\ Q_{\pm} &\approx& \bar{Q}~\int dQ~dp~N_{\pm}(Q,p)~\approx~ \bar{Q}~N_+~,
\eea
where $E(\bar{Q},\bar{p})$ is the average energy of one Q-ball 
with the average charge $\bar{Q}$ and average momentum $\bar{p}$. Now the energy charge ratio of Q-balls (anti-Q-balls), $x_+=E_+/mQ_+~~ (x_-=E_-/mQ_-)$, determines the average momentum of Q-balls
\be{momentum1}
\bar{p} \approx m~\bar{Q}~\sqrt{x_{\pm}^2-1}\gsim 0.1~m~\bar{Q}~,
\ee
when $x_{\pm}\gsim 1.01$. The average velocity will then be
$<v>=\bar{p}/E(\bar{Q},\bar{p})\gsim 0.1$. 
If Q-balls were non-relativistic, it would cause a restriction
$x_{\pm}-1<0.01$, which is not natural, as will be seen later.
Thus the main bulk of Q-balls, except the largest ones, 
may be expected to be relativistic.

Since most Q-balls are relativistic it is not surprising that the 
reaction rate turns out to be much larger than the Hubble rate. We find
\be{rate1}
\Gamma = \frac{N_{tot}}{V}<\sigma v> \approx 
{4\pi \rho\bar{p} R^2\over E(\bar{Q},\bar{p})^2}~,
\ee
where we have approximated the cross section by the geometrical 
cross section with $R$ the radius of the Q-ball. The cross section
actually depends on the relative phase difference between the
colliding Q-balls \cite{mv2} as is largest when the phases are
aligned; however, the order of magnitude is that of the
geometrical cross section. 

When $x\gg 1$, the charge and the momenta of Q-balls and anti-Q-balls are almost the same. 
The reaction rate \eq{rate1} is at its maximum for $\bar{p}=m\bar{Q}$
corresponding to $x_{\pm}\approx \sqrt{2}$, which is about the same as the equilibrium value.
The average charge is $\bar{Q}=q_+V_1$
where $q_+$ is the charge density inside a Q-ball and $V_1$ it's volume.
Thus we can rewrite (\ref{rate1}) as $\Gamma={3V_1\rho/ 2m\bar{Q}R}={3\rho/ 2m q_+R}$.
In the dense-packing limit, $\rho\approx 2mq_+x_+\approx 2\sqrt{2}\,mq_+$,
and hence
\be{rate2}
\Gamma \approx \frac{1}{R}~.
\ee

We approximate the radius of Q-balls by $R\approx |K|^{-1/2}m^{-1}$
\eq{potEMcD} and compare this to the Hubble rate $H\sim mt_0/t$ to find that
$\Gamma>H$ when
\be{rate3}
t \gsim \frac{t_0}{|K|^{1/2}}
\ee
Since typically $|K|\sim 0.01-0.1$ \cite{ejm},
the reaction rate will be larger than the Hubble rate 
from the very beginning,
and a thermal equilibrium can be expected to be reached. 

Guided by these considerations, let us now find the equilibrium 
distributions for Q-balls in the gravity-mediated case in two and three
dimensions. The one-particle partition function reads
\be{part1}
Z_1 = g\integral{V_D}{} \frac{d^D{x}~d^D{p}}{(2\pi)^D}
\integral{-\infty}{\infty} dQ~e^{-\beta E+\mu Q},
\ee
where $\mu$ is the chemical potential related to the charge of Q-balls. 
In $D$ dimensions we may integrate \eq{part1} to obtain a distribution
function with respect to charge \be{part2}
Z_1 = \integral{-\infty}{\infty}dQ~Z_1(Q) = 2gV_D\beta\left(\frac{m}{2\pi\beta}
\right)^{\frac{D+1}{2}}\integral{-\infty}{\infty}dQ~|Q|^{\frac{D+1}{2}}~
e^{\mu Q}~K_{\frac{D+1}{2}}(\beta m|Q|)~,
\ee
where $K_n(z)$ is the modified Bessel function. Then we may integrate \eq{part2} 
to obtain an expression that can be split into positive and negative charge
parts: \bea{part5}
Z_1 &=& Z_1^- + Z_1^+ \nn
Z_1^{\pm} &=& \frac{2gV_D\beta m^{D+1}(D+1)!}{\pi^{\frac{D}{2}}\Gamma(\frac{D}{2}+2)} 
\left[\frac{F(D+2,\frac{D}{2}+1;\frac{D}{2}+2:-\frac{\beta m\pm\mu}{\beta
m\mp\mu})}{(\beta m\mp\mu)^{D+2}} \right]~,
\eea
where $F(a,b;c:x)$ is the hypergeometric function.
For $D=2$ and $D=3$ these expressions simplify to
\bea{part6}
Z_1^{\pm} &=& \frac{gV_2}{2\pi\beta^2}\frac{2\beta m\mp\mu}{(\beta m\mp\mu)^2},~D=2~; \\
Z_1^{\pm} &=& \frac{gV_3}{\pi^2\beta^3}\left[\pm
\frac{\mu(5\beta^2m^2-2\mu^2)}{2(\beta^2m^2-\mu^2)^2}+
\frac{3\beta^4m^4}{(\beta^2m^2-\mu^2)^{\frac{5}{2}}}\arctan\sqrt{\frac{\beta
m\pm\mu}{\beta m\mp\mu}}\right],~D=3. \nn \label{part61} \eea
The total partition function is the sum of these two terms:
\bea{part7}
Z_1 &=& \frac{2gV_2\beta m^3}{\pi(\beta^2m^2-\mu^2)^2}, \qquad D=2~; \\
Z_1 &=& \frac{3gV_3\beta m^4}{2\pi(\beta^2m^2-\mu^2)^{\frac{5}{2}}}, \qquad D=3. \label{part71}
\eea
The grand canonical partition function is given, as usual, by
\be{grandp}
Z_G=\sum_{N=0}^{\infty}{e^{\bar{\mu} N}Z_1^N\over N!}=e^{zZ_1}
\ee
($z=e^{\bar\mu}$) with the energy, charge and number of Q-balls given
respectively as \be{thermo}
E_{tot} = -\frac{\partial\log Z_G}{\partial\beta};~~Q_{tot} = 
\frac{\partial\log Z_G}{\partial\mu};~~ N_{tot} = \frac{\partial\log
Z_G}{\partial\bar{\mu}}. \ee
Using \eq{part2} in \eq{thermo} the total energy of the Q-ball system is
obtained
\be{endist1}
E_{tot} = mN_{tot}~\integral{-\infty}{\infty}dQ~f_E(Q)~,
\ee
where
\be{endist2}
f_E(Q) = \frac{2}{6-D}~(2\pi)^{-\frac{D-1}{2}}~\left[1-\left(\frac{\mu}{\beta m}\right)^2
\right]^{\frac{D}{2}+1}~(\beta m|Q|)^{\frac{D+3}{2}}~e^{\mu
Q}~K_{\frac{D+3}{2}}(\beta m|Q|)~
\ee
is the energy distribution function per Q-ball scaled with the mass $m$.
The average energies of Q-balls and anti-Q-balls are then found from \eqs{part6}{part61} to be
\bea{energy1}
E_{\pm} &=& \frac{N_{tot}}{2\beta^4m^3}
\frac{(\beta m\pm\mu)^2(3\beta^2m^2\mp 3\beta m\mu+\mu^2)}{\beta m\mp\mu},
~~D=2~; \\ E_{\pm} &=&
\frac{2N_{tot}}{\pi\beta}\Big[\pm\frac{\mu}{2\beta^4m^4}
\frac{6\beta^4m^4-5\beta^2m^2\mu^2+2\mu^4}{\sqrt{\beta^2m^2-\mu^2}} \nn  &+&
\frac{4\beta^2m^2+\mu^2}{\beta^2m^2-\mu^2}
\arctan\sqrt{\frac{\beta m\pm\mu}{\beta m\mp\mu}}\Big], ~~D=3.\nn & & \label{energy11}
\eea
The total energy is, again, the sum of positive and negative parts:
\bea{energy2}
E_{tot} = \frac{N_{tot}}{\beta}\frac{3\beta^2m^2+\mu^2}{\beta^2m^2-\mu^2}, ~~D=2~; \\ E_{tot} = 
\frac{N_{tot}}{\beta}\frac{4\beta^2m^2+\mu^2}{\beta^2m^2-\mu^2}, ~~D=3.
\eea
Similar calculations can also be repeated for charge and
particle number. We get the total charge 
\be{chadist1}
Q_{tot} = N_{tot}\integral{-\infty}{\infty}dQ~\sign(Q)~f_Q(Q)~,
\ee
where
\be{chadist2}
f_Q(Q) = \frac{2~(2\pi)^{-\frac{D-1}{2}}}{6-D}~\left[1-\left(\frac{\mu}{\beta m}\right)^2
\right]^{\frac{D}{2}+1}~(\beta m|Q|)^{\frac{D+3}{2}}~e^{\mu
Q}~K_{\frac{D+1}{2}}(\beta m|Q|)~
 \ee
is the charge distribution function per Q-ball.
The positive and negative parts of the total charge are:
\bea{charge1}
& &Q_{\pm} = \frac{N_{tot}}{4\beta^3m^3}
\frac{(\beta m\pm\mu)^2(3\beta m\mp\mu)}{\beta m\mp\mu}, ~~D=2~; \\
& &Q_{\pm} = \frac{N_{tot}}{3\pi}\left[
\frac{8\beta^4m^4+9\beta^2m^2\mu^2-2\mu^4}{\beta^4m^4\sqrt{\beta^2m^2-\mu^2}}
\pm\frac{30\mu}{\beta^2m^2-\mu^2}\arctan\sqrt{\frac{\beta m\pm\mu}{\beta
m\mp\mu}}\right], ~~D=3~.\nn & & \label{charge11}
\eea
The charges are defined to be positive so that the total charge is $Q_{tot}=Q_+-Q_-$. Thus
\bea{charge2}
Q_{tot} = N_{tot}\frac{4\mu}{\beta^2m^2-\mu^2}, \qquad D=2~; \\
Q_{tot} = N_{tot}\frac{5\mu}{\beta^2m^2-\mu^2}, \qquad D=3~. \label{charge21}
\eea
Moreover, the relative number distribution function of Q-balls fulfilling
condition \be{number3}
\beta m \integral{-\infty}{\infty}dQ~f_N(Q)=1
\ee
reads now
\be{number31}
f_N(Q) = \frac{2(2\pi)^{-\frac{D-1}2}}{6-D}~\left[1-\left(\frac{\mu}{\beta
m}\right)^2\right]^{\frac{D}{2}+1}~ (\beta m|Q|)^{\frac{D+1}{2}}~e^{\mu
Q}~K_{\frac{D+1}{2}}(\beta m|Q|)~.
 \ee
Finally, the number of positive and negative charge Q-balls is then found to
be \bea{number2}
N_{\pm} &=& \frac{N_{tot}}{4\beta^3m^3}
(2\beta m\mp\mu)(\beta m\pm\mu)^2,\ D=2\; ;\\
N_{\pm} &=& \frac{N_{tot}}{\pi}\left [
\pm\frac{\mu(5\beta^2m^2-2\mu^2)}{3\beta^4m^4} \sqrt{\beta^2m^2-\mu^2}+
2\arctan\sqrt{\frac{\beta m\pm\mu}{\beta m\mp\mu}}\right ],\ D=3\;.\nn \eea

We have plotted the charge and number distribution functions for 
$-10\leq\beta mQ\leq 10$ in $D=2$ and $D=3$ for 
$\mu=0,~0.1\beta m$ in Figs. \ref{chadist} 
and \ref{numdist}. The energy distribution function has a
behaviour which is  qualitative similar to  
the number distribution function. The total numbers of 
Q-balls and anti-Q-balls in $D=2$ and $D=3$ are plotted in 
Fig. \ref{numdens} for $1\leq x\leq 30$. When $x\geq 30$ there 
are almost equal numbers of Q-balls and anti-Q-balls.

\begin{figure}[ht]
\leavevmode \centering \vspace*{7cm}
 \includegraphics{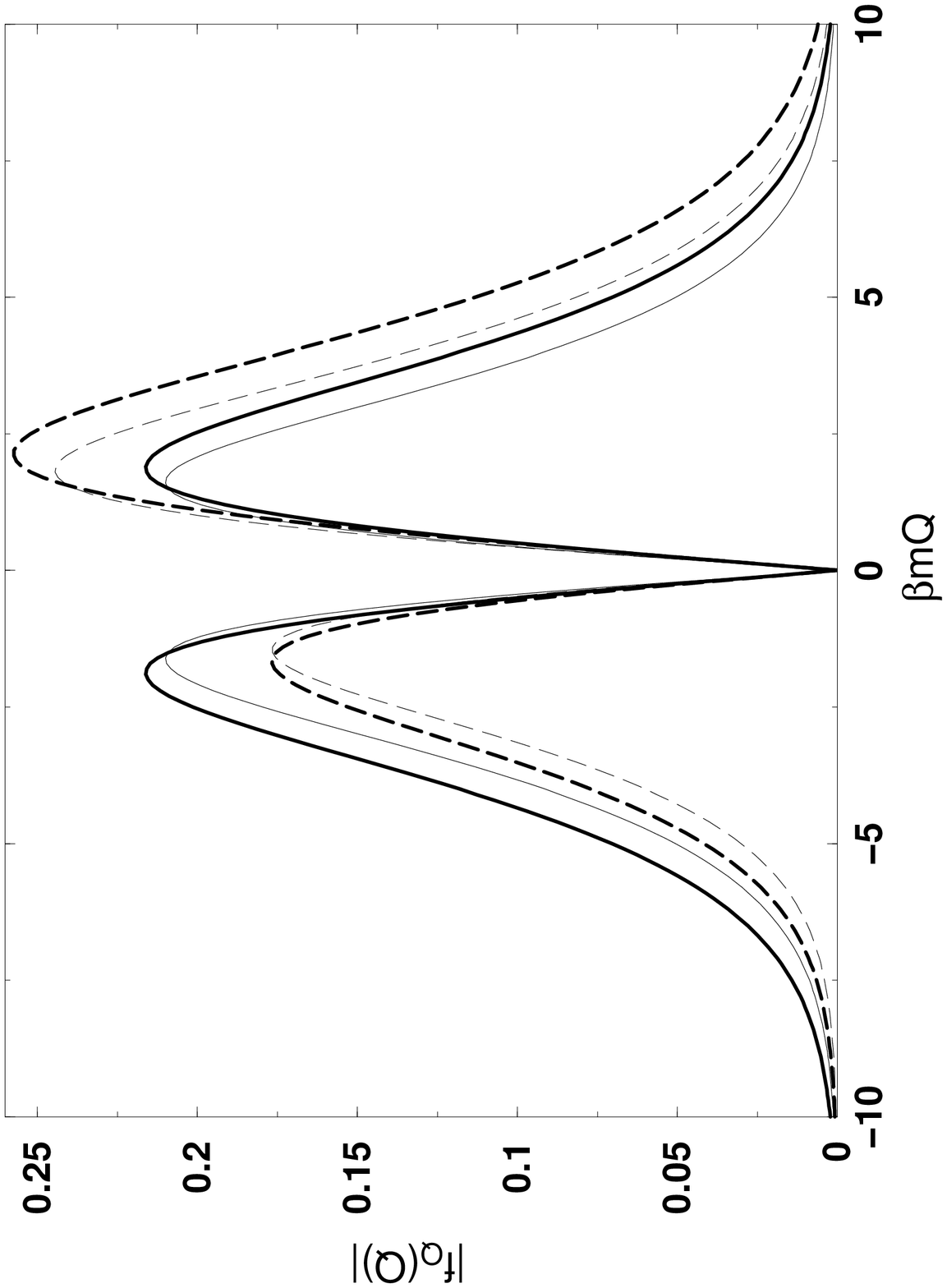}
 \caption{Charge distributions of Q-balls for $\mu=0$ (solid lines) and $\mu=0.1\beta m$ 
(dashed lines) in $D=2$ (thin lines) and $D=3$ (thick lines).}
\label{chadist} \vspace*{12cm}  \includegraphics{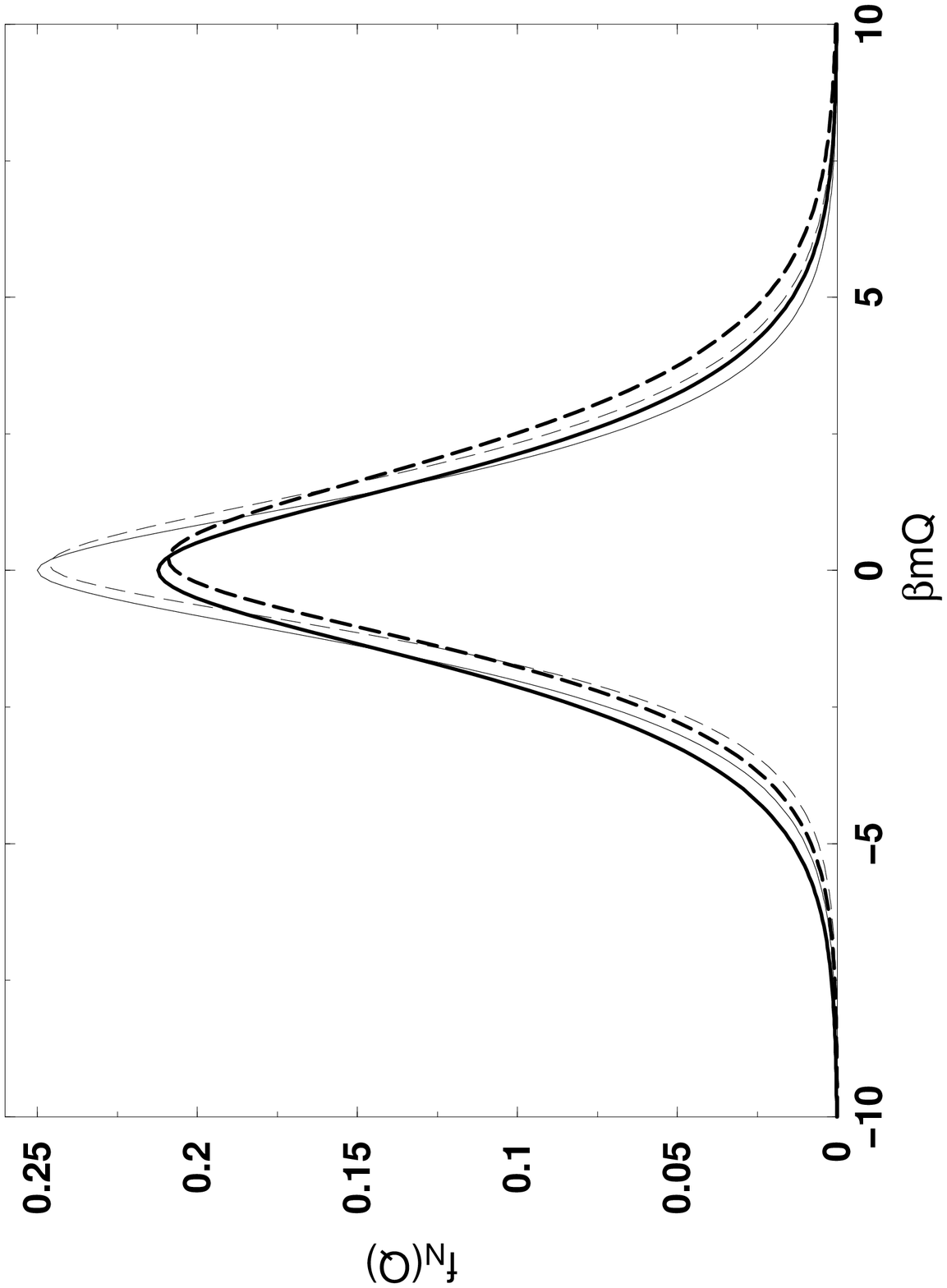}  \caption{Number distributions of
Q-balls for $\mu=0$ (solid lines) and $\mu=0.1\beta m$ (dashed lines) in $D=2$
(thin lines) and $D=3$ (thick lines).} \label{numdist} \end{figure} 
\begin{figure}[ht]
\leavevmode \centering \vspace*{7cm}
 \includegraphics{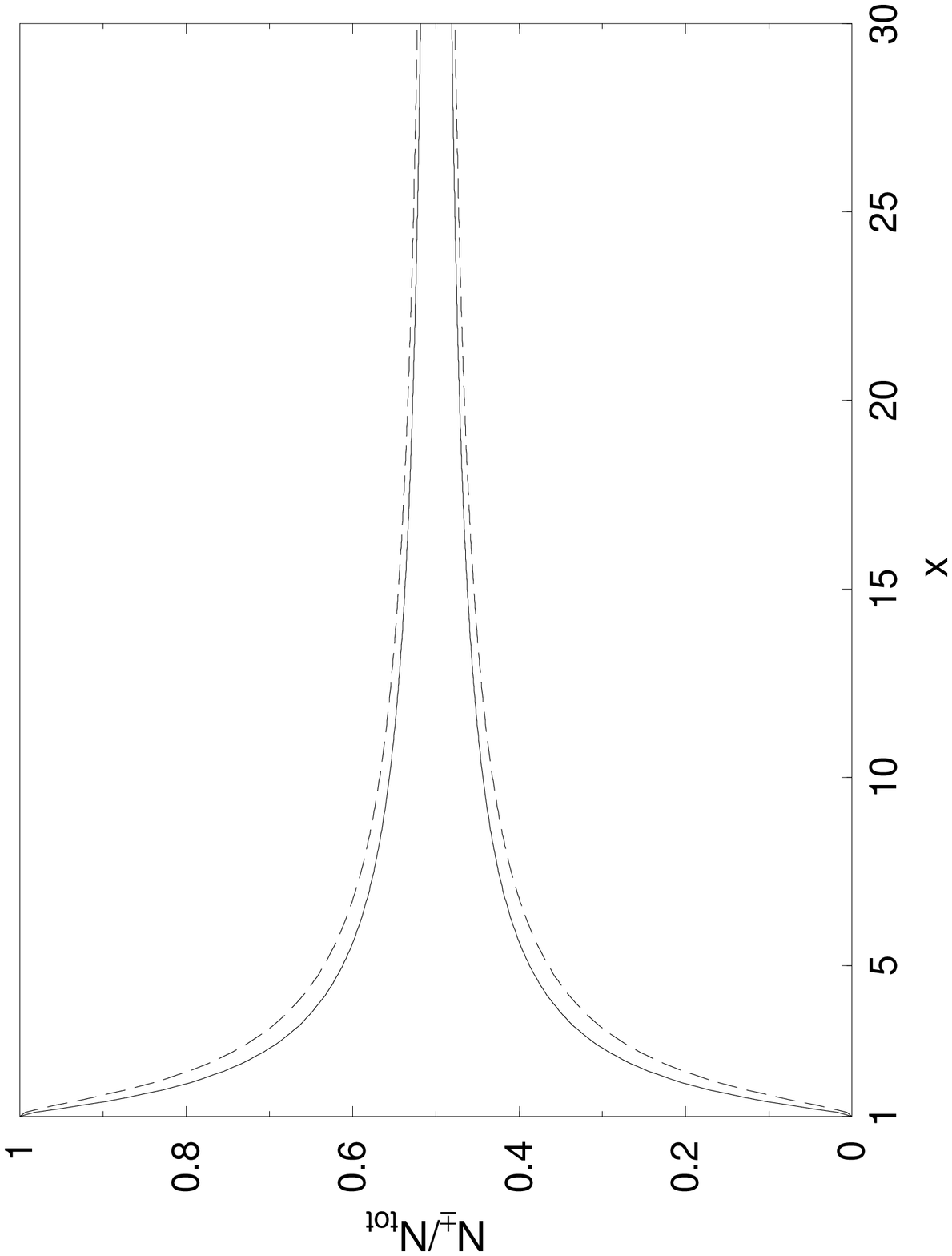}
 \caption{Number densities $N_+$ ($N_-$) of Q-balls (anti-Q-balls), curves above $0.5$ 
(curves below $0.5$), in $D=2$ (solid lines) and $D=3$ (dashed lines).}
\label{numdens} \vspace*{12cm}  \includegraphics{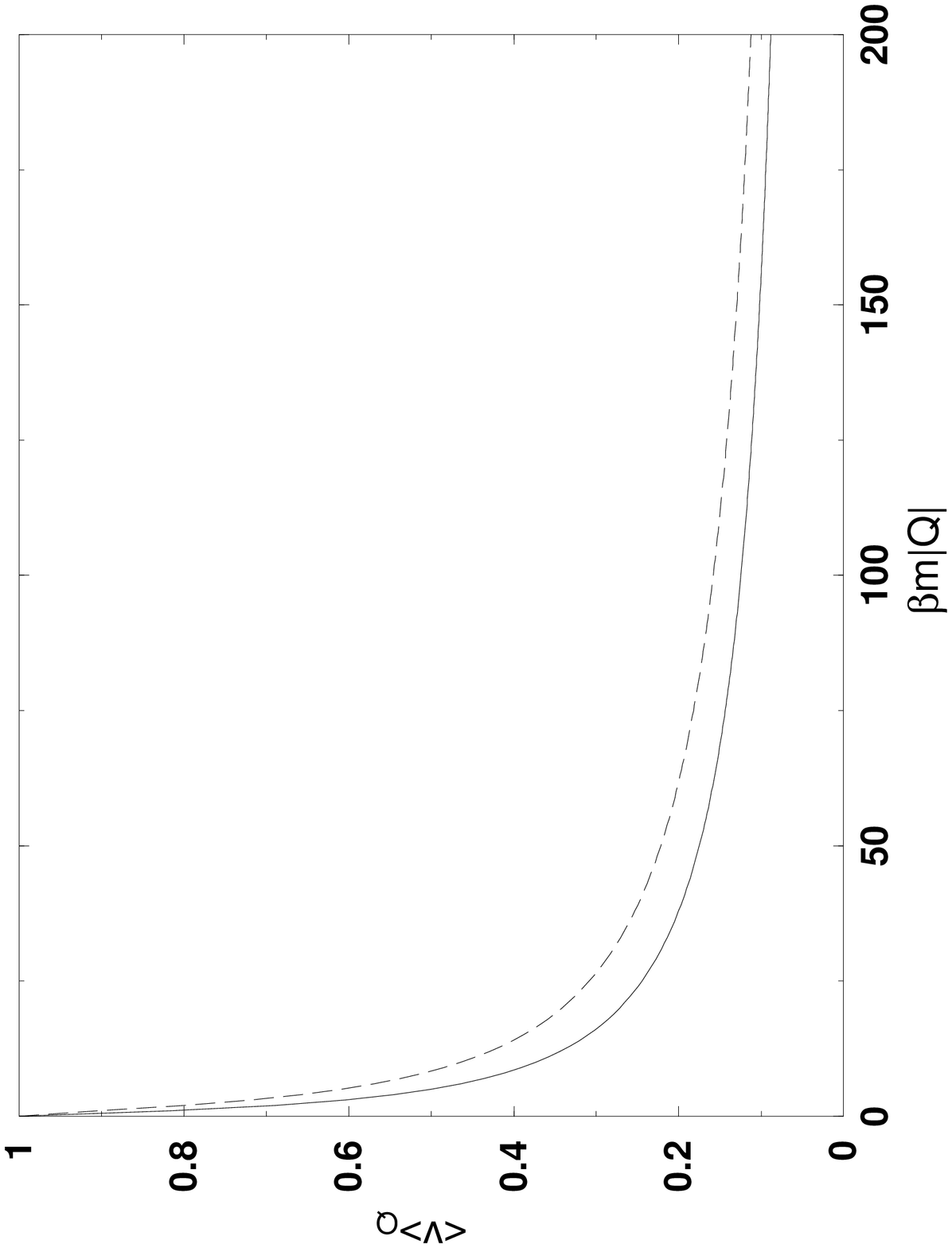}  \caption{Velocity distributions in
$D=2$ (solid line) and $D=3$ (dashed line).} \label{veldist} \end{figure}

When total energy and charge are fixed, 
the chemical potential $\mu$ can be found to read
\bea{muu2}
\mu = \beta m\left(2x-\sqrt{4x^2-3}\right), \qquad D=2~;\\
\mu = \frac{\beta m}{2}\left(5x-\sqrt{25x^2-16}\right), \qquad D=3~. \label{muu3}
\eea

Using one-particle partition function \eq{part1} we can also calculate the average 
velocity of Q-balls to be
\be{velocity1}
<v> = \frac{1}{Z_1} g\integral{V_D}{}\frac{d^D{x}~d^D{p}}{(2\pi)^D}\integral
{-\infty}{\infty} dQ~\frac{p}{E}~e^{-\beta E + \mu Q} =
\frac{1}{Z_1}\integral{-\infty}{\infty}dQ~Z_1(Q)~<v>_Q~, \ee
from which we get, using \eq{part2}, the velocity distribution function, 
which is plotted in Fig. \ref{veldist},
\be{velocity2}
<v>_Q~ = \frac{\Gamma(\frac{D+1}{2})}{\Gamma(\frac{D}{2})}~\sqrt{\frac{2}{\beta m|Q|}}~
\frac{K_{\frac{D}{2}}(\beta m|Q|)}{K_{\frac{D+1}{2}}(\beta m|Q|)}~.
\ee
The average velocity gives again the hypergeometric function and finally we have
\bea{velocity3}
<v> = \frac{\pi}{4\beta m}\sqrt{\beta^2m^2-\mu^2}~,\qquad D=2~; \\
<v> = \frac{8}{3\pi\beta m}\sqrt{\beta^2m^2-\mu^2}~,\qquad D=3.
\eea

Assuming $x\gg 1$, we see that the chemical potential will almost vanish because 
$\mu\sim\cO (1/x)$. This in turn implies that the amounts of positive and
negative charges, \eq{charge1}, will be much larger than the total charge,
\eq{charge2}, which is of the order of $\cO (\mu)$. Note that the average
velocity is relativistic: $\pi/4$ for $D=2$ and $8/3\pi$ for $D=3$ but the
velocity distribution shows that for small Q-balls the velocity is close to one
whereas for large Q-balls we get non-relativistic velocities. (Thus  the
qualitative behaviour in these two cases is the same, as can be seen in
Fig. \ref{numdens}.) Moreover,  
in the large $x$ limit (vanishing $\mu$ limit)
the energy to charge ratio of Q-balls and
anti-Q-balls, $x_{\pm}=E_{\pm}/mQ_{\pm}$, is easily calculated. Using
\eqs{energy1}{energy11} and \eqs{charge1}{charge11} we obtain
$x_{\pm}\approx 2$, regardless of dimension.
 
Using the equilibrium distributions one can verify
that the reaction rate is larger than the Hubble rate so that the assumptions
in this Section are self-consistent. The real situation is naturally expected
to be much more complex, but as we will see in the next Section, from
numerical simulations one extracts distributions which are very close to the
thermal ones discussed here. 
\section{Numerical simulations}  \subsection{Preliminaries}

Because analytical considerations of the previous Section
indicate that Q-ball distributions in two and three dimensions behave 
qualitatively in the same way, it is reasonable to expect that two
dimensional simulations will also shed light on the more realistic three
dimensional case. The great advantage is of course, that much less CPU time is
needed for the simulation. Therefore we have simulated Q-ball formation
numerically on a $2+1$ dimensional lattice. The largest lattice size used in
simulations was $100\times 100$ (simulations on smaller lattices were run to
study lattice size effects).  We take as the initial condition a uniform
AD-condensate with an arbitrary phase $\w$, 
\be{initcond} 
\Phi=\phi_0 \textrm{e}^{i\w t}+\delta\Phi 
\ee 
with uniformly distributed random noise $\delta\Phi$ added to the amplitude 
and phase. The amplitude ratio of the noise and the condensate 
field, $|\delta\Phi|/|\phi_0|$, is here $\cO(10^{-13})$. The phase of 
the AD-condensate varies in the range $\w=10^{0},\ 
10^{-1},...,10^{-5}$ which corresponds to $x\sim 1,...,10^5$. 
The initial amplitude of the AD-field is set to $\phi_0=10^9\GeV$.
This is smaller than the actual condensate amplitude in d=4 flat
direction \cite{bbb1} but should be large enough for the
simulation to encompass all the qualitative features of condensate
fragmentation.

The equation of motion of the AD-condensate is 
\be{adeqm} 
\ddot\Phi+3H\dot\Phi-{1\over a^2}\nabla^2\Phi+ 
m^2\Phi[1+{K\over\ln {10}}+K\log({|\Phi|^2\over 
M^2})]-g H^2|\Phi|+ {3\lambda^2\over\MP^2}|\Phi|^4\Phi=0, 
\ee 
where $M$ is a large mass scale, $a$ is the scale factor of the 
universe and $H$ is the Hubble parameter, $H={\dot a/ a}$.

For the simulations
the field was decomposed into real and imaginary parts,
$\phi={1\over\sqrt{2}}(\phi_1+i\phi_2)$. We also rescale 
field and
space-time according to
\be{recalings}
\varphi={\phi\over m},\ \ h={H\over m},\ \ \tau=mt,\ \ \xi=mx.
\ee

The parameter values chosen for the simulations were
$m=10^2\GeV,\ K=-0.1$ and $\lambda={1\over 2}$ (the $g H^2$-term can be
omitted since $g H^2\ll m^2$). The universe is assumed to be matter
dominated so that $a=a_0t^{2\over 3}$ and $H={2\over 3t}$. The
spatial lattice unit was $\Delta\xi=0.1$ and the temporal unit
$\Delta\tau=0.9\Delta\xi$. The calculations were done in a
comoving volume so that the physical size of the lattice
increases with time. The initial time is $\tau_0=100$ and
$a=0.1\times\tau^{2\over 3}$. The simulations were run up to
$\sim 2\times 10^6$ time steps.

\subsection{Outline of numerical results}

We have studied six different cases,
$\w=10^0,10^{-1},...,10^{-5}$ ($x\sim 1,...,10^5$) on differently sized lattices. 
(Note that even though $x$ may cosmologically be as large as $10^8$ the main features
of the Q-ball formation process are the same for any $x\gg 1$. Increasing the initial
energy in the condensate only increases the relaxation time of the condensate and
hence CPU time required for the simulation.)
In all of the six cases the qualitative description of the evolution 
of the AD-condensate at the beginning of the simulation is 
similar. First the charge density of the condensate decreases 
uniformly in the box due to the expansion of the universe (the 
charge in the comoving volume is constant throughout the 
simulation). No large fluctuations are visible yet at this time 
and the fluctuation spectrum corresponds to the white noise 
present in the initial conditions. As time progresses a growing 
mode can be seen to develop. White noise is still present but 
the growing mode soon starts to dominate. This process continues 
until lumps of positive charge develop. 

The further evolution of the AD-condensate depends on the initial 
energy-to-charge ratio $x$ of the condensate and hence on the 
value of $\w$ in the initial conditions. If $x=1$, \ie the 
energy-to-charge ratio is equal to that of a Q-ball, the evolution 
of the AD-condensate continues in a similar fashion. As the 
universe continues to expand the lumps of positive charge slowly develop
into Q-balls while their spatial distribution effectively freezes in the
comoving volume. No negatively charged Q-balls, anti-Q-balls, are formed in
this case. After going non-linear, the lumps just evolve into Q-balls and
finally freeze due to the expansion of the universe. This is
in fact exactly the
case  studied in \cite{kasuya2}, where the formation of Q-balls was followed
in a three dimensional simulation. No anti-Q-balls were however observed 
since $x=1$.

On the other hand, if $\w$ is smaller than one so that $x\gg 1$, 
as seems more natural, the Q-ball 
formation process is much more complicated. After the positively 
charged lumps have formed, expanded linearly and then
developed non-linearly, the extra energy
stored in them causes the lumps to fragment as they evolve into Q-balls.
In this process a large number of negatively charged Q-balls
is formed. The total charge in the negative and positive
Q-balls is approximately equal so that the initial charge in the condensate is
in fact negligible compared to the amount of charge and anti charge created. 

As the universe continues expanding, the Q-ball and anti-Q-ball 
number density distributions freeze as they tend towards the
equilibrium distributions discussed in the previous Section.

\subsection{Evolution of the AD-condensate}

We have plotted the charge density as a 3D-plot and as a contour
plot in the comoving volume at different time intervals in the case $\w=10^{-5}$ in
Figures \ref{form1}-\ref{form12}. One should note that the z-scale of
the 3D-plots and the gray scale coloring of the contour plots varies 
from Figure to Figure\footnote{Colour coded versions of the figures,
where more details can be seen, can be found at
www.utu.fi/\char 126 tuomul/qballs/}.

\begin{figure}[ht]
\leavevmode \centering \vspace*{7cm}
\includegraphics{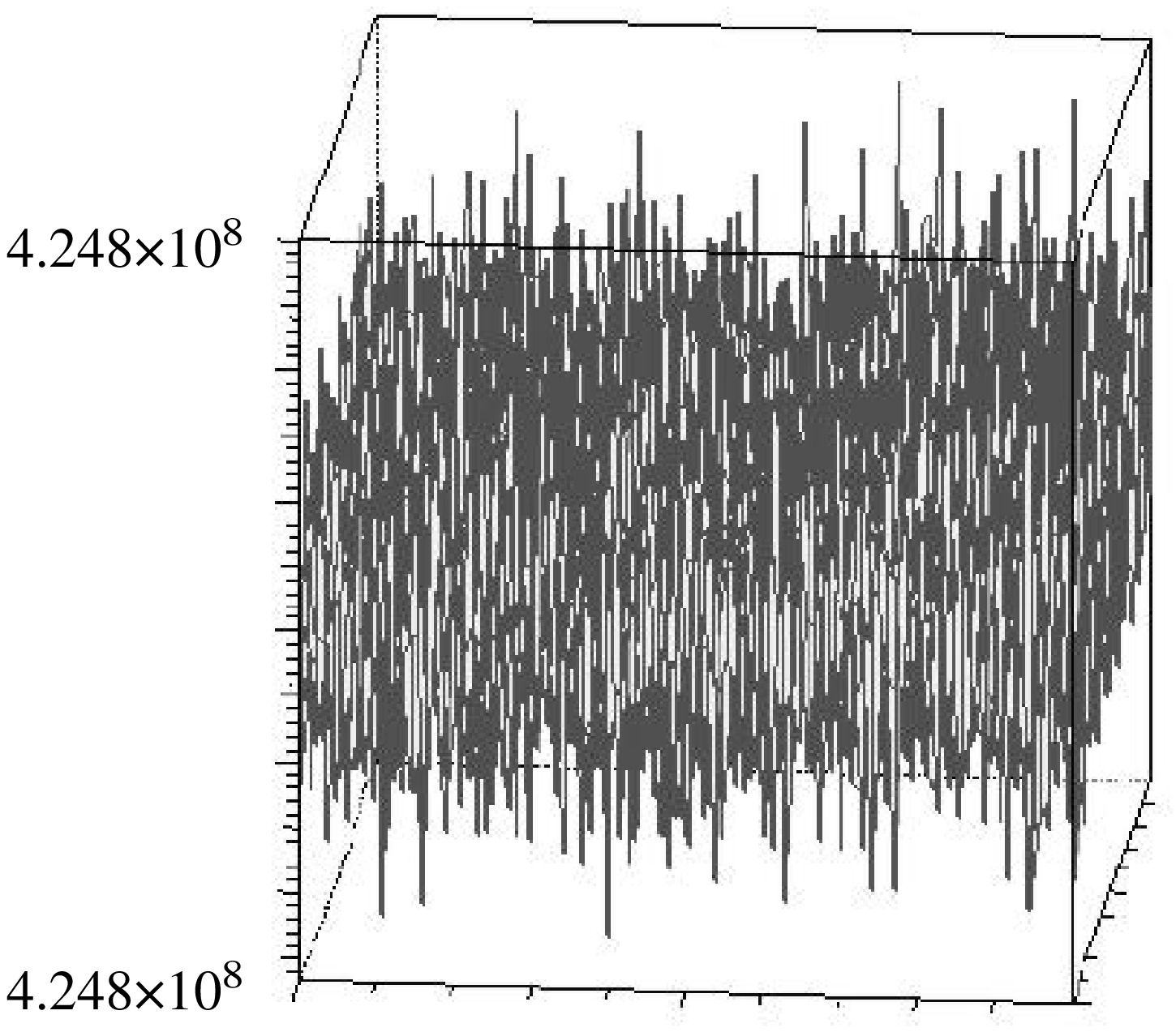}
\includegraphics{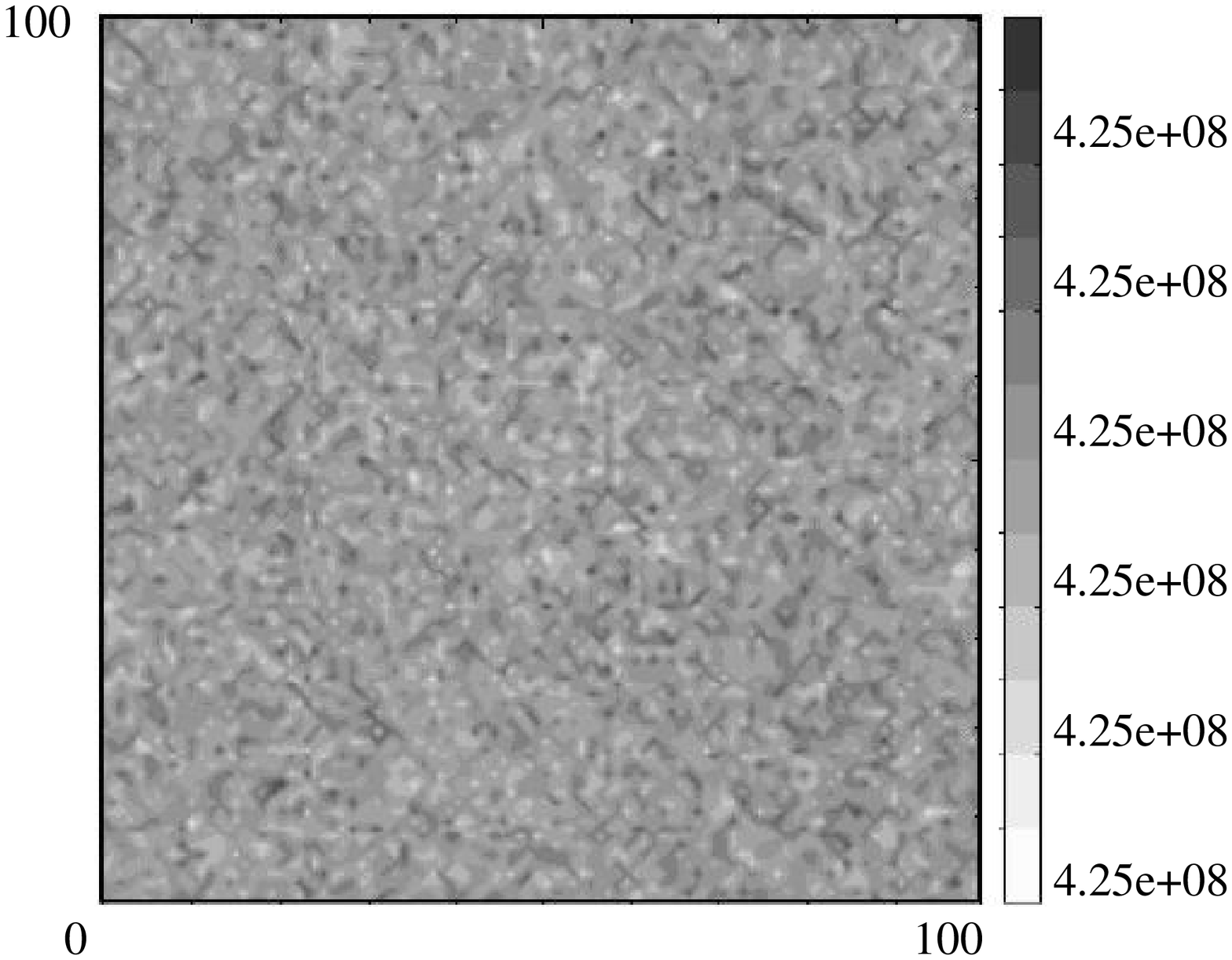}
\caption{$\tau=90$} \label{form1} \vspace*{7cm}
\includegraphics{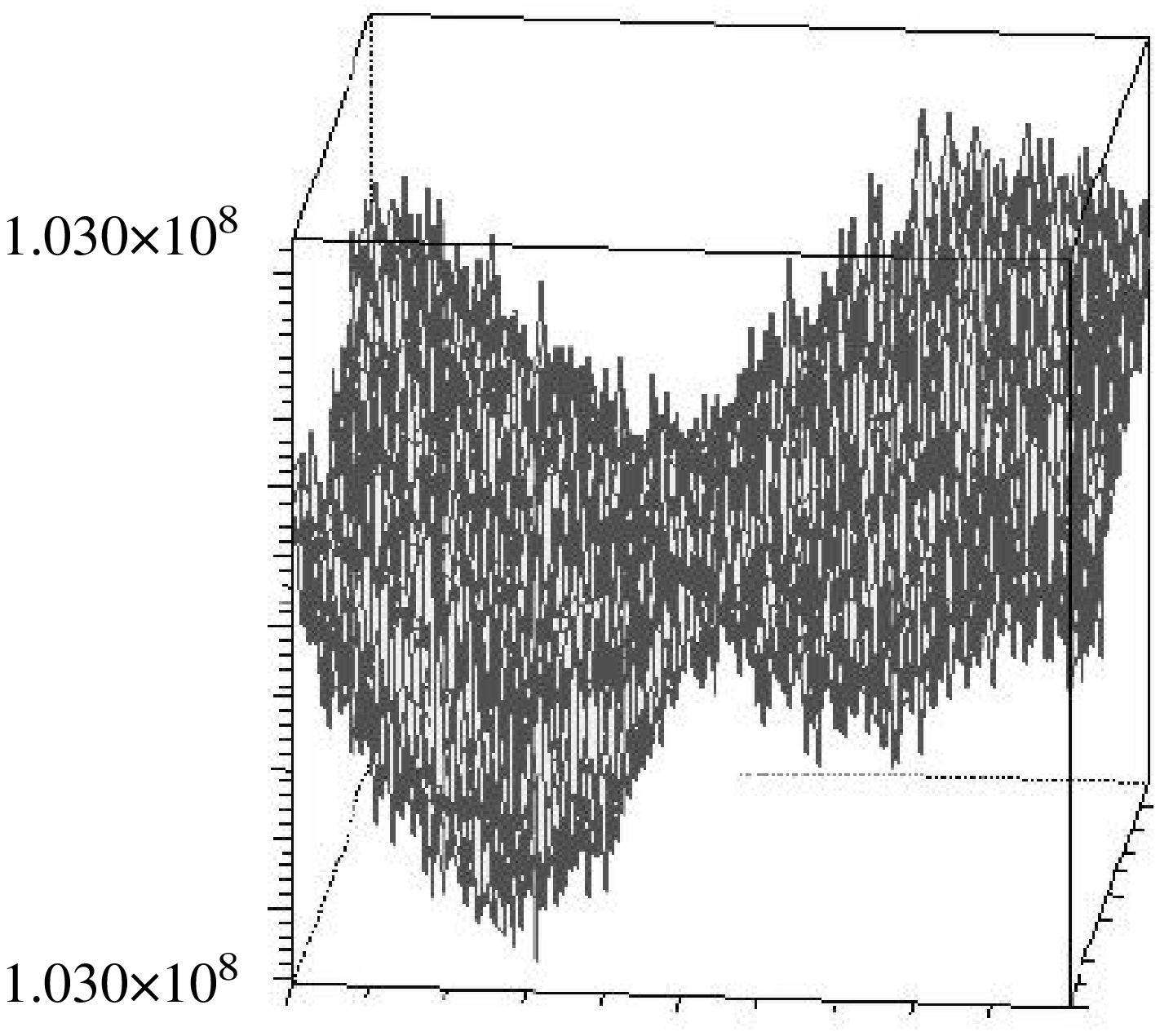}
\includegraphics{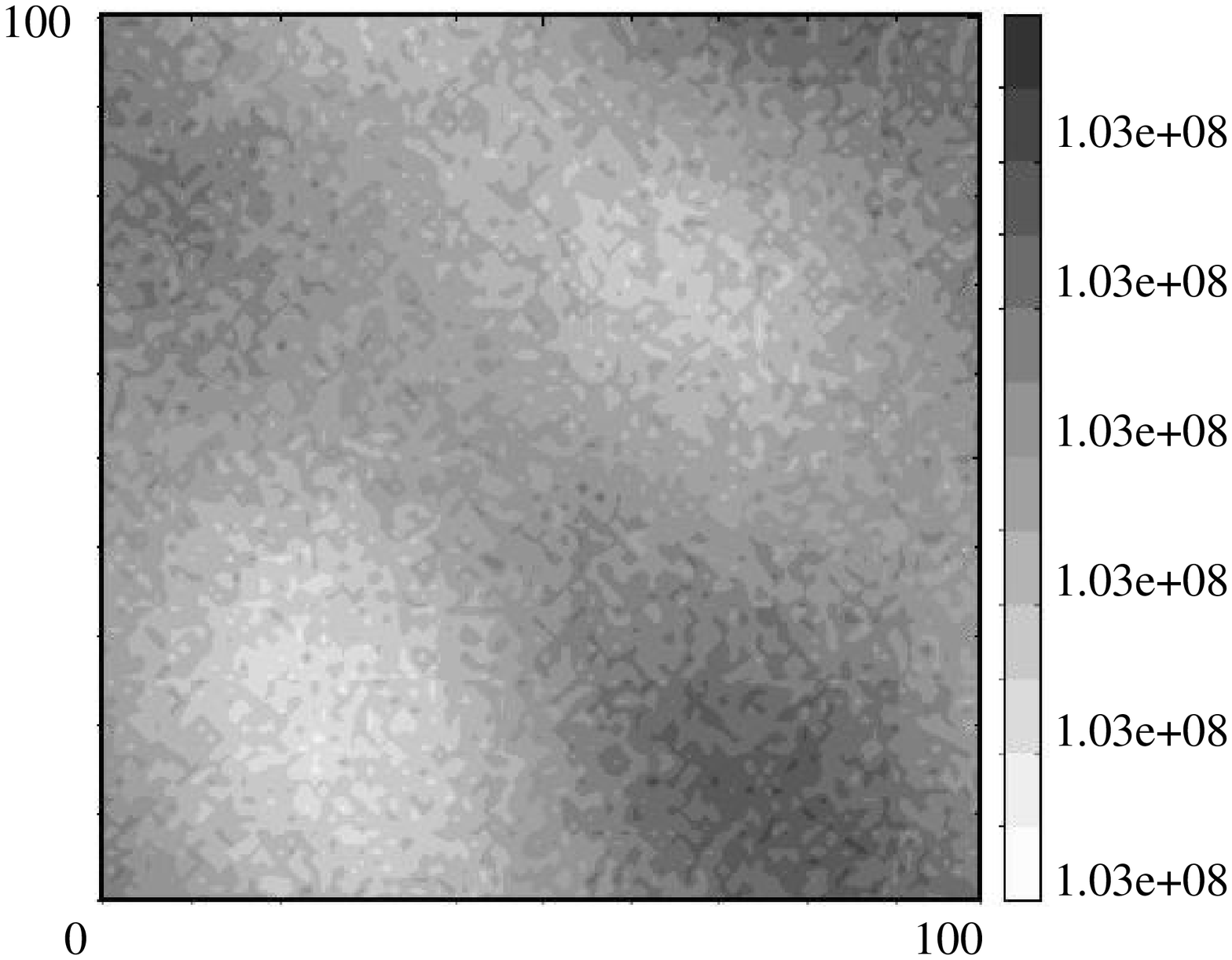}
\caption{$\tau=450$} \label{form2} \vspace*{7cm}
\includegraphics{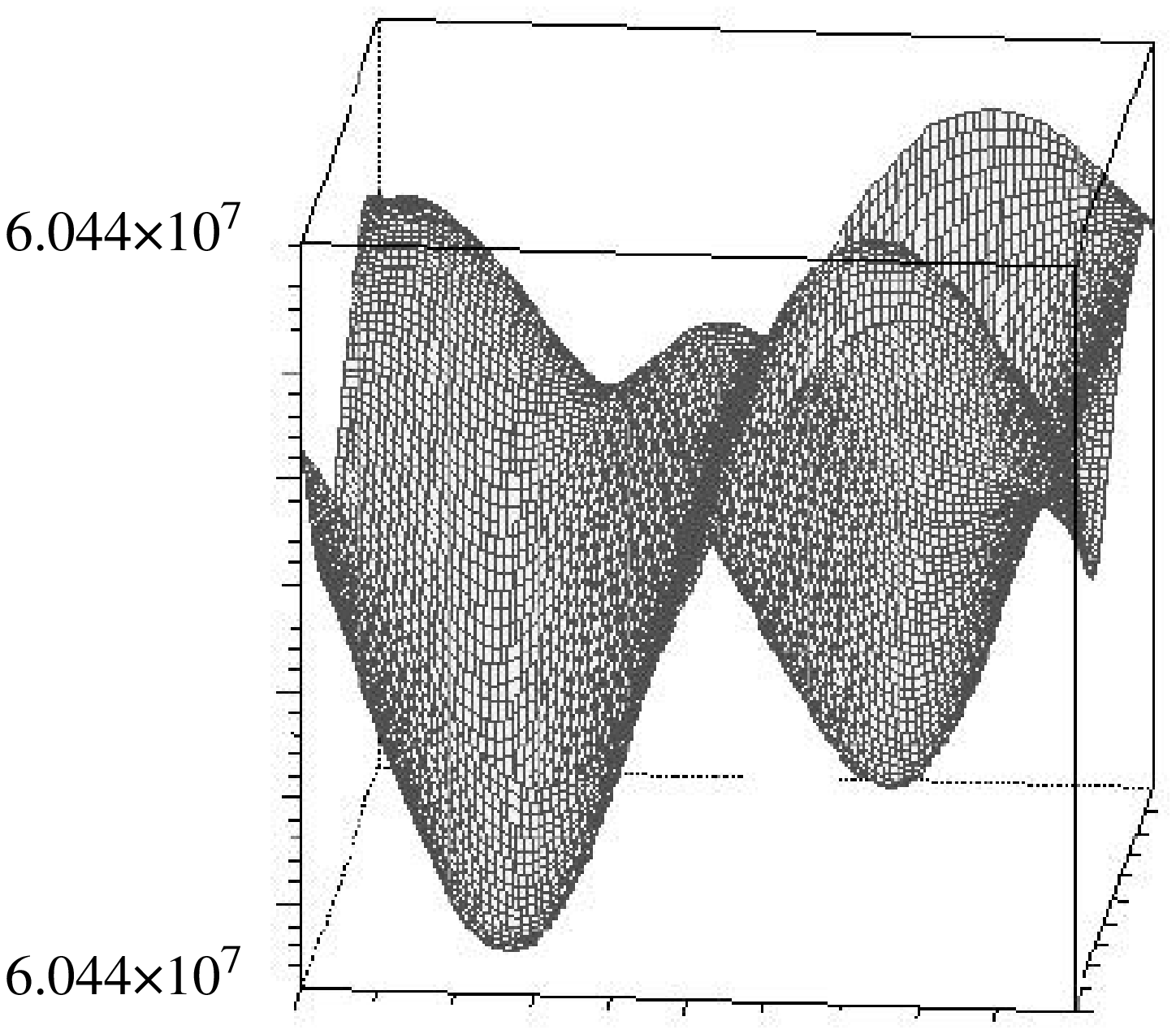}
\includegraphics{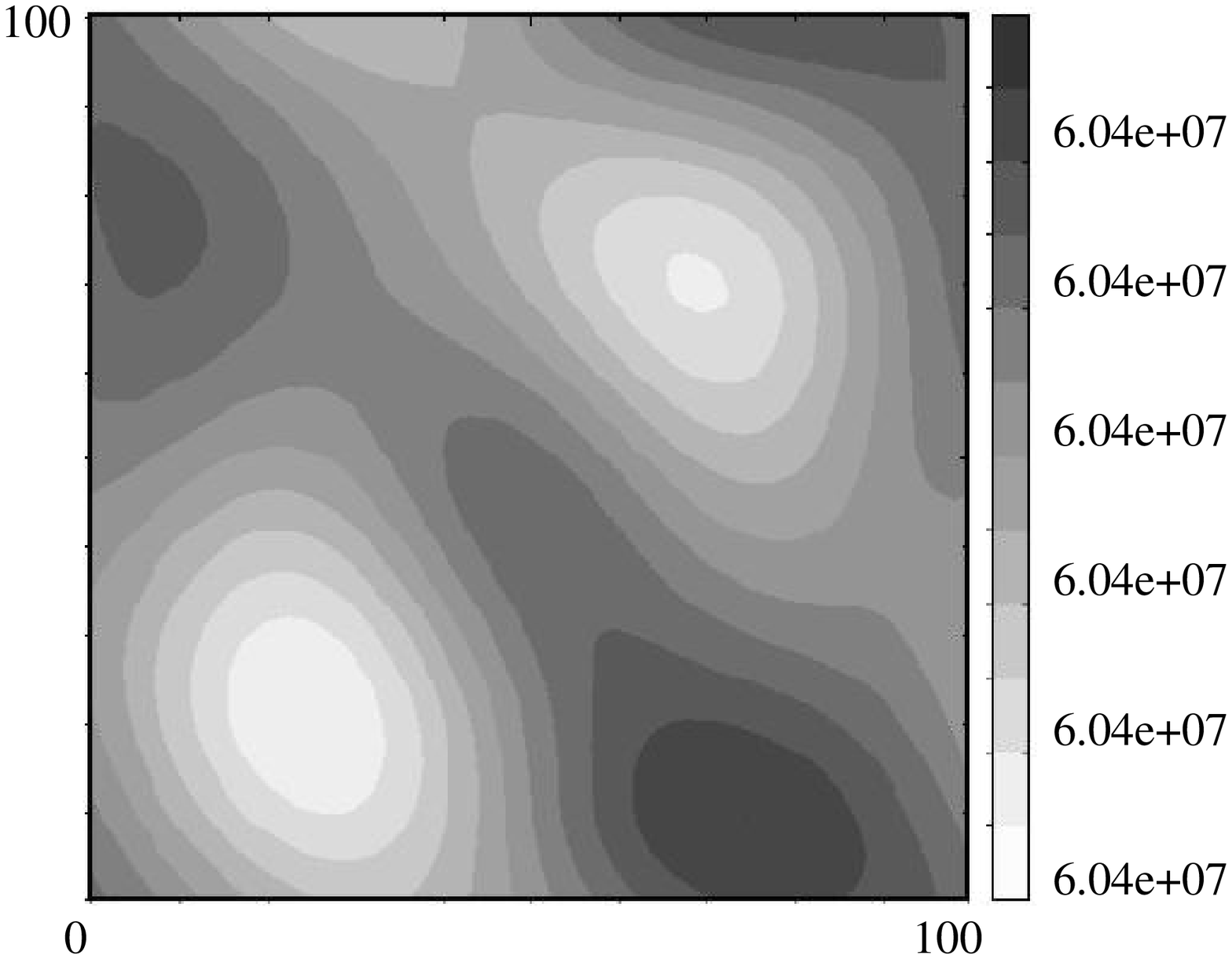}
\caption{$\tau=720$}\label{form3}
\end{figure}

\begin{figure}[ht]
\leavevmode \centering \vspace*{7cm}
\includegraphics{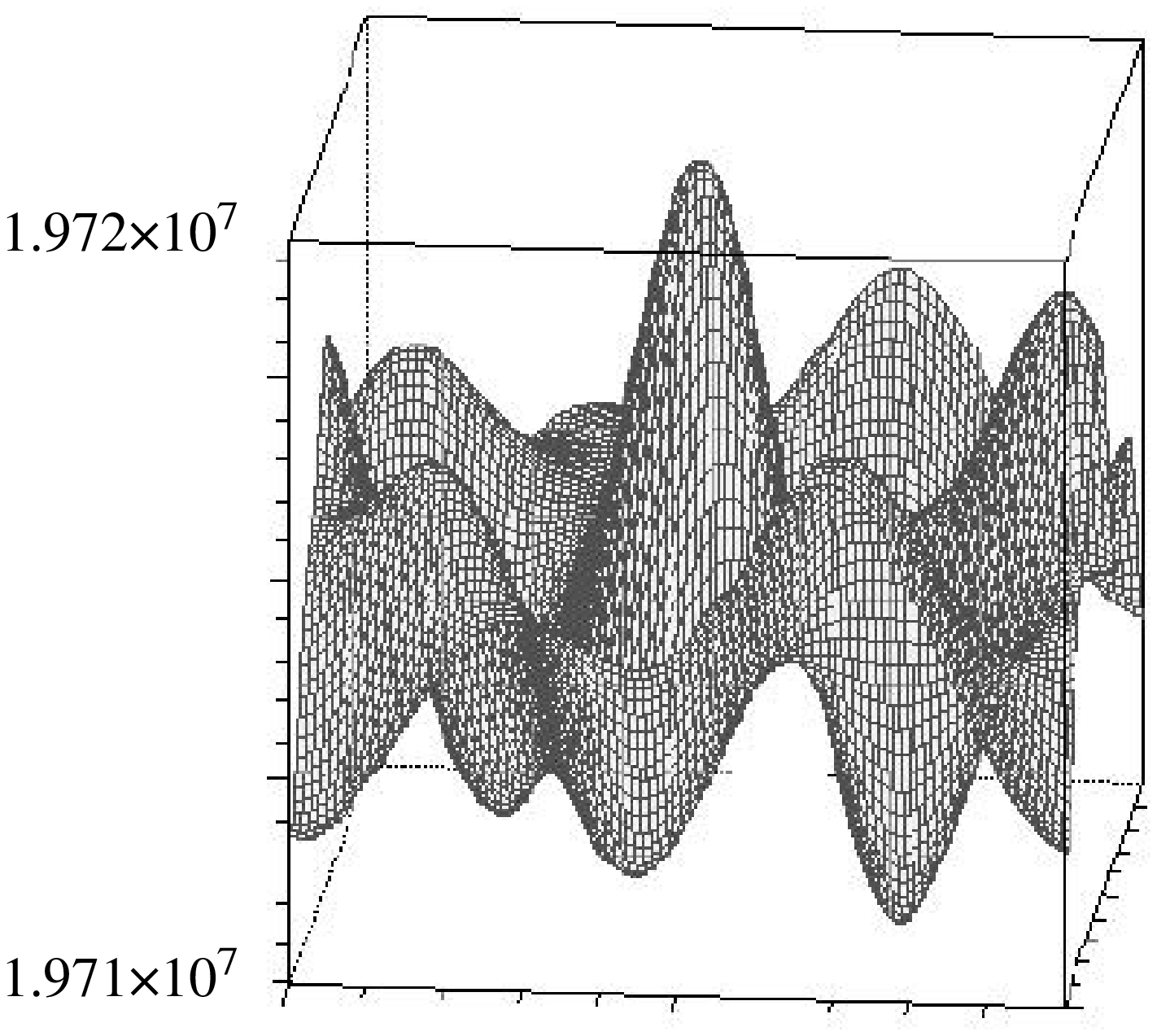}
\includegraphics{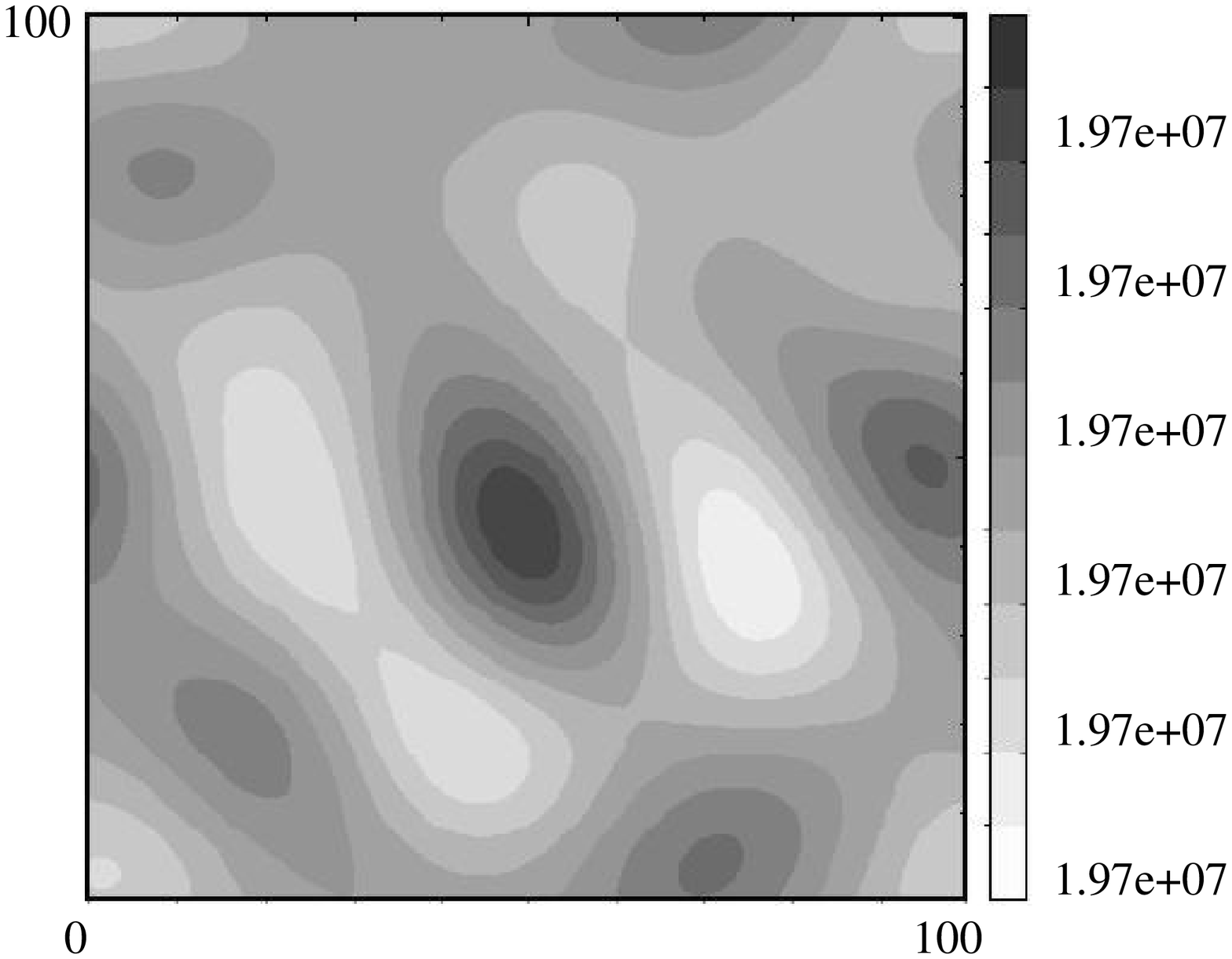}
\caption{$\tau=1800$} \label{form4} \vspace*{7cm}
\includegraphics{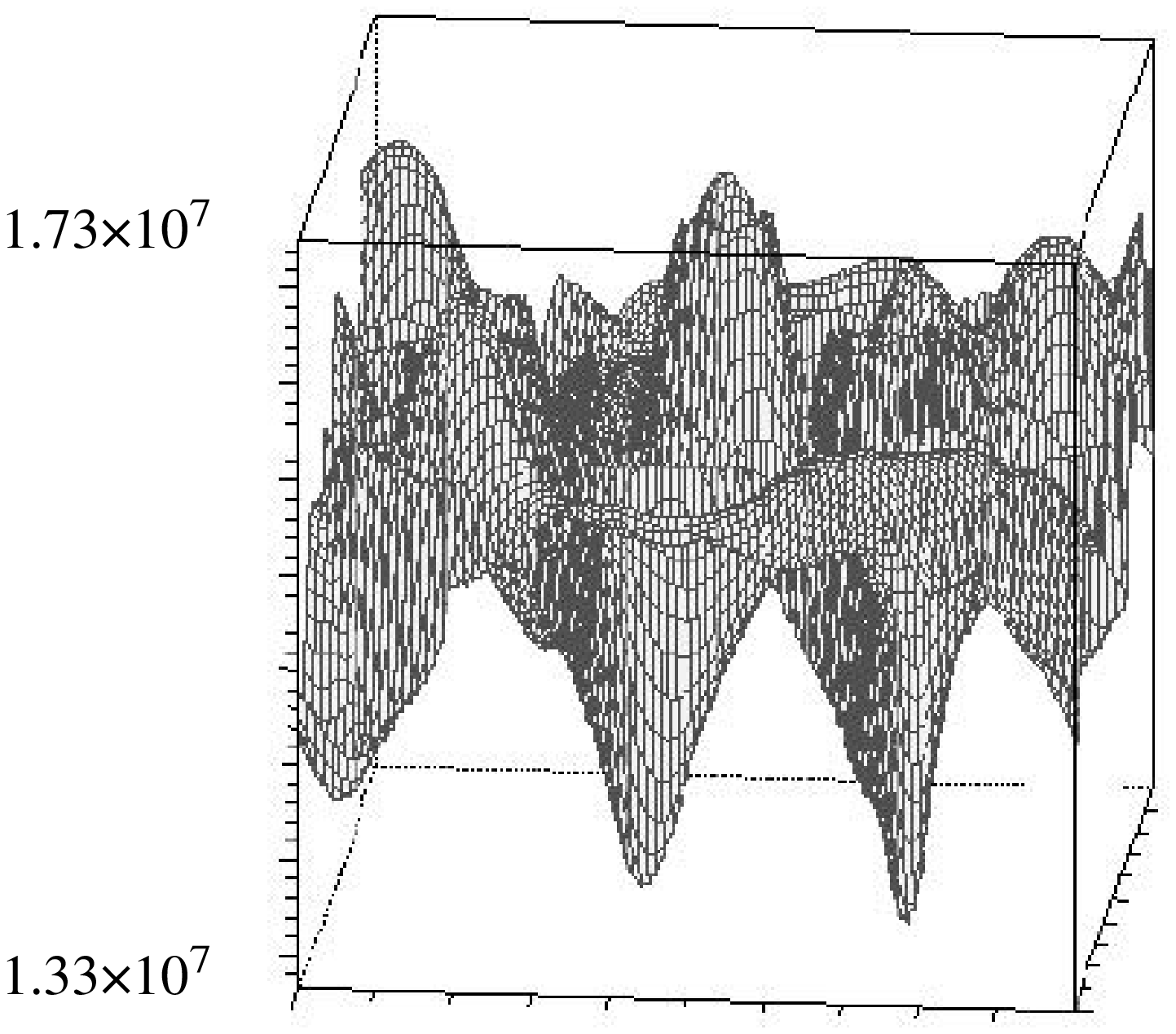}
\includegraphics{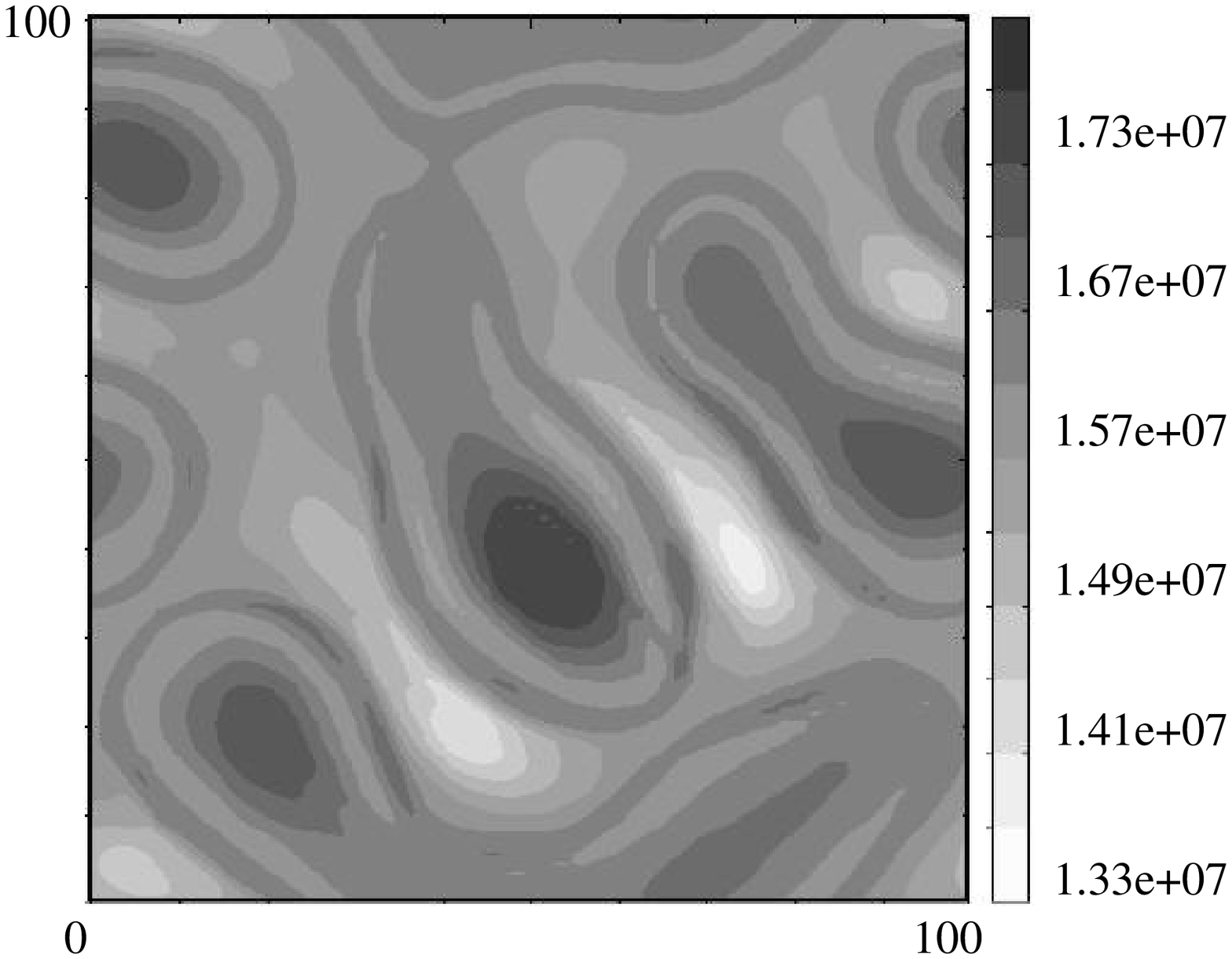}
\caption{$\tau=2160$}\label{form5}  \vspace*{7cm}
\includegraphics{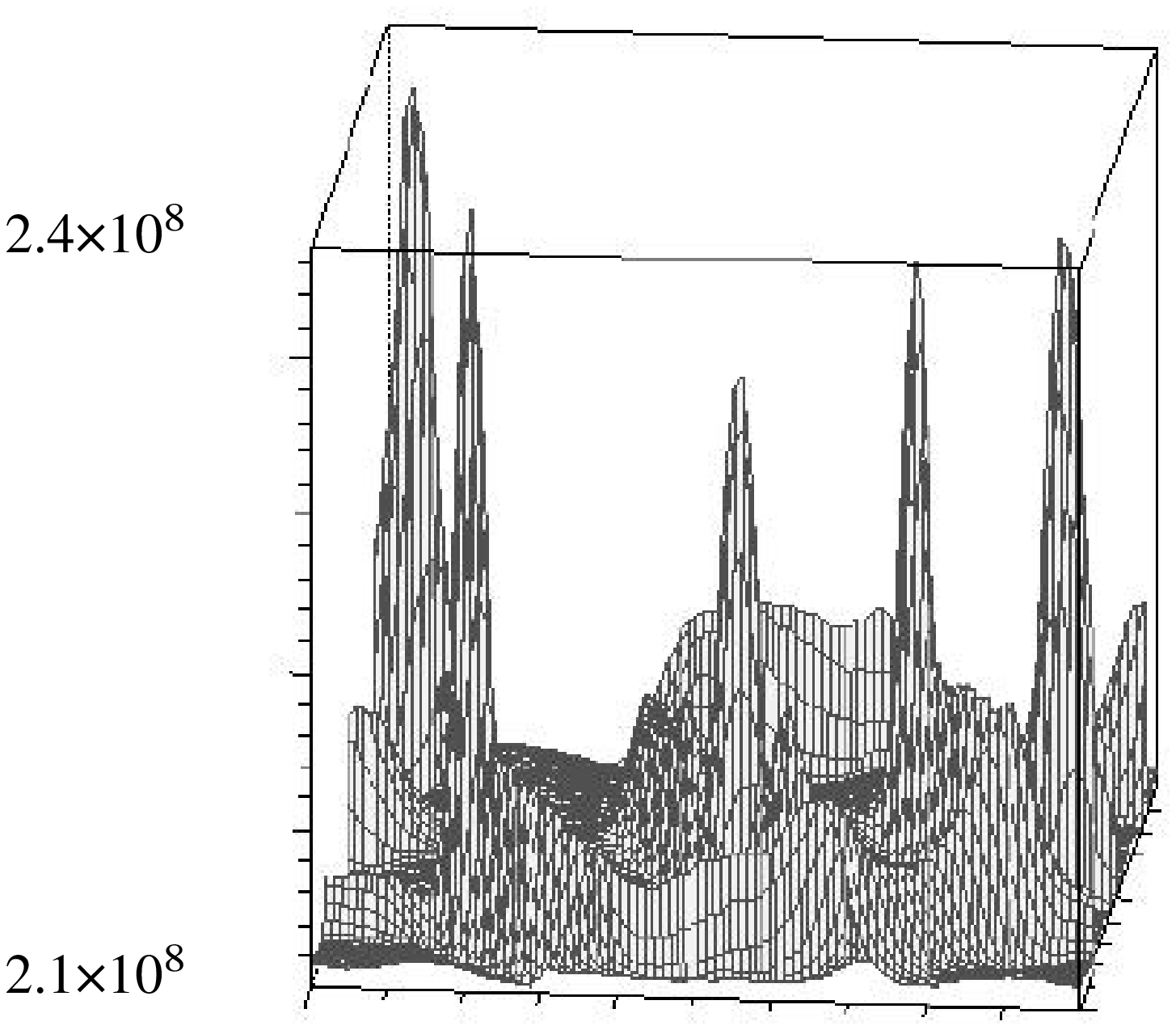}
\includegraphics{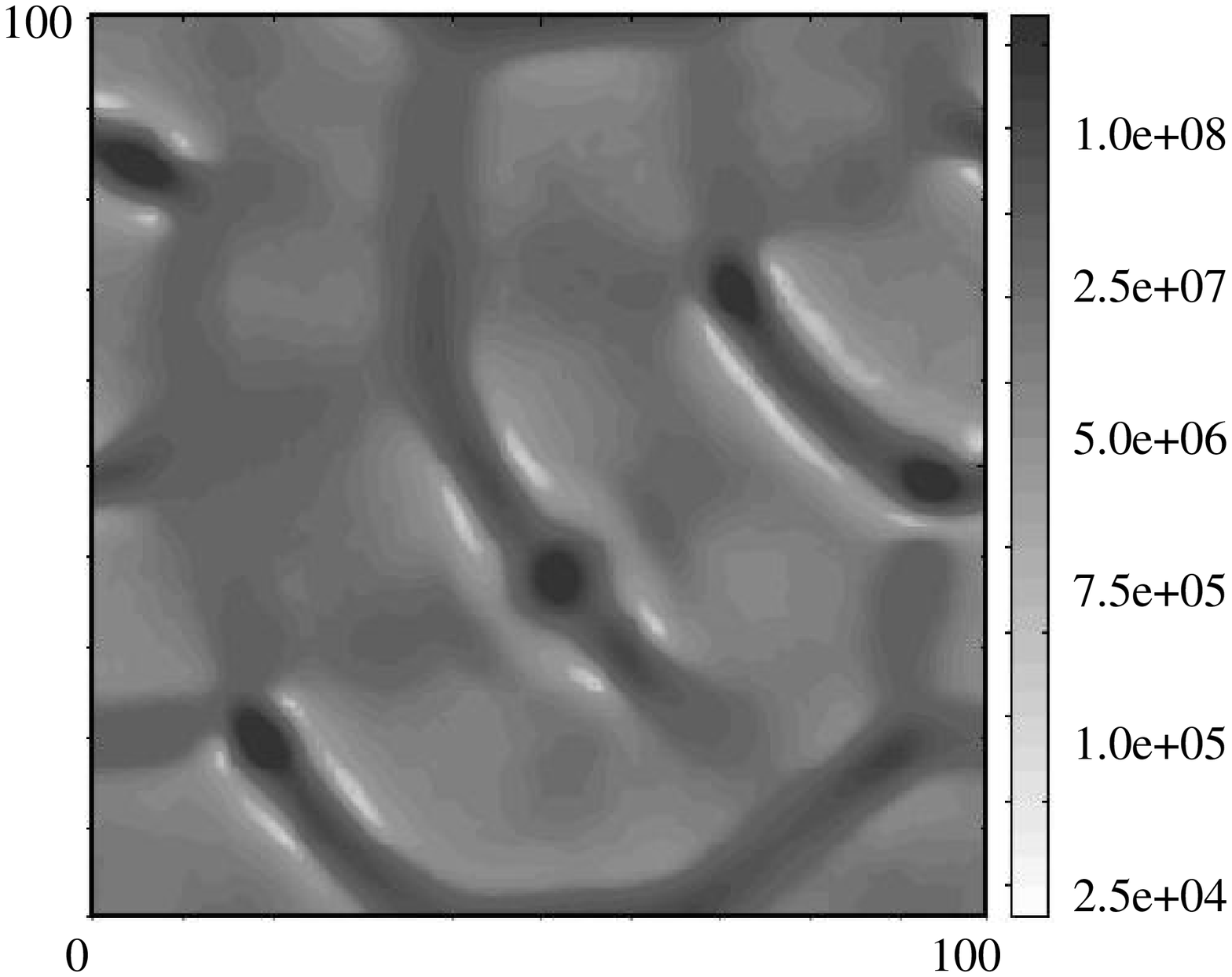}
\caption{$\tau=2430$} \label{form6} 
\end{figure}

\begin{figure}[ht]
\leavevmode \centering \vspace*{7cm}
\includegraphics{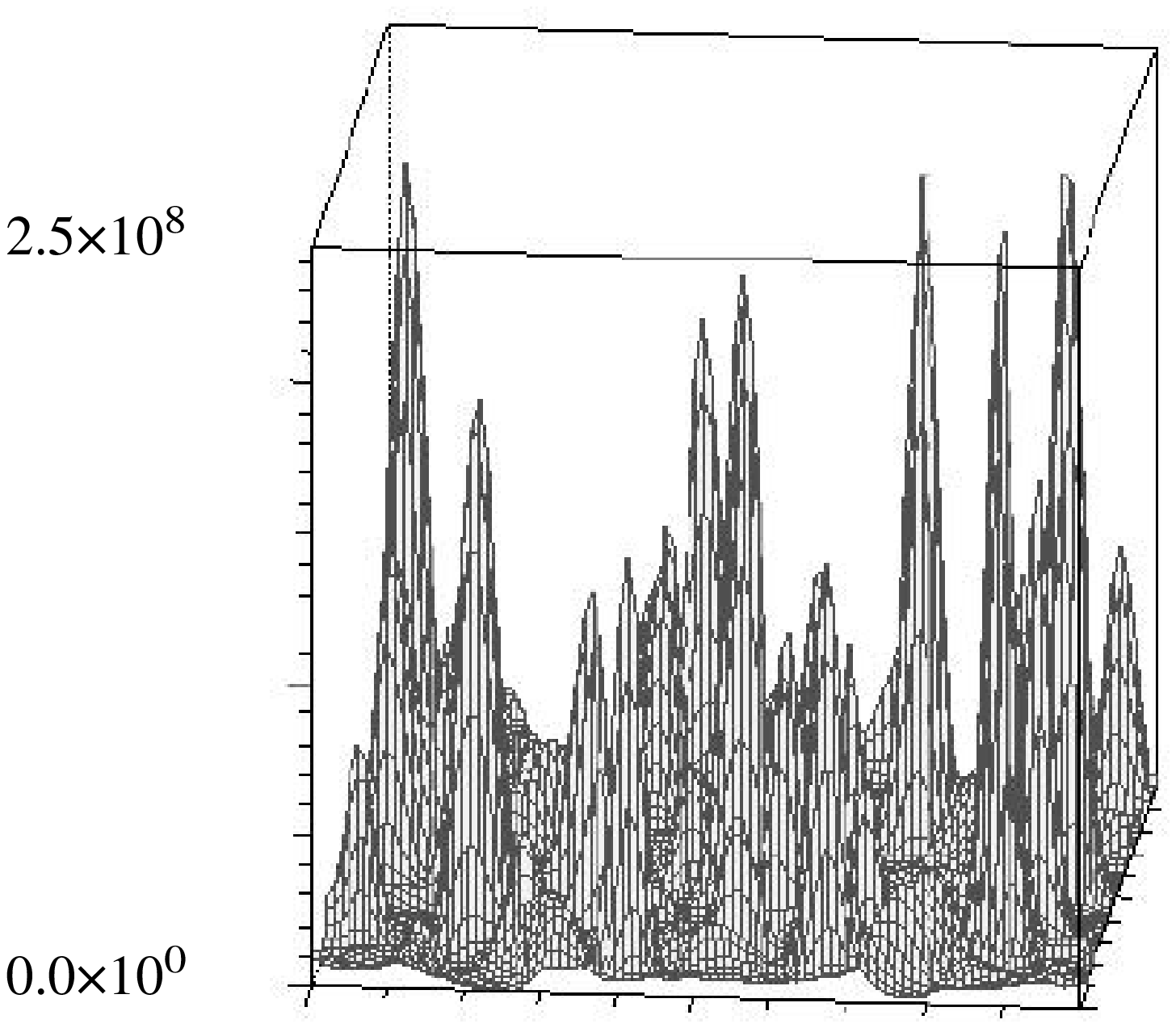}
\includegraphics{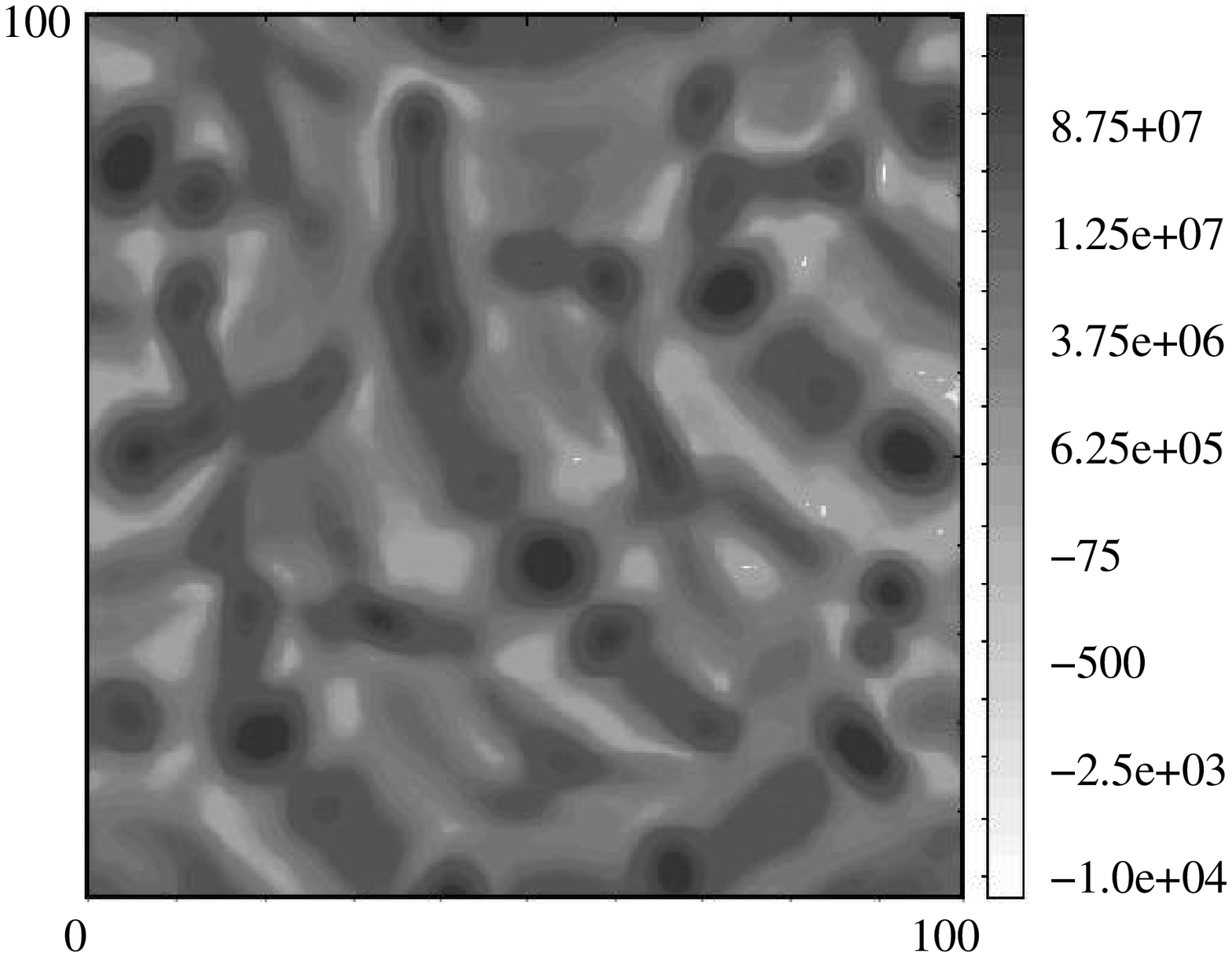}
\caption{$\tau=2520$}\label{form7} \vspace*{7cm}
\includegraphics{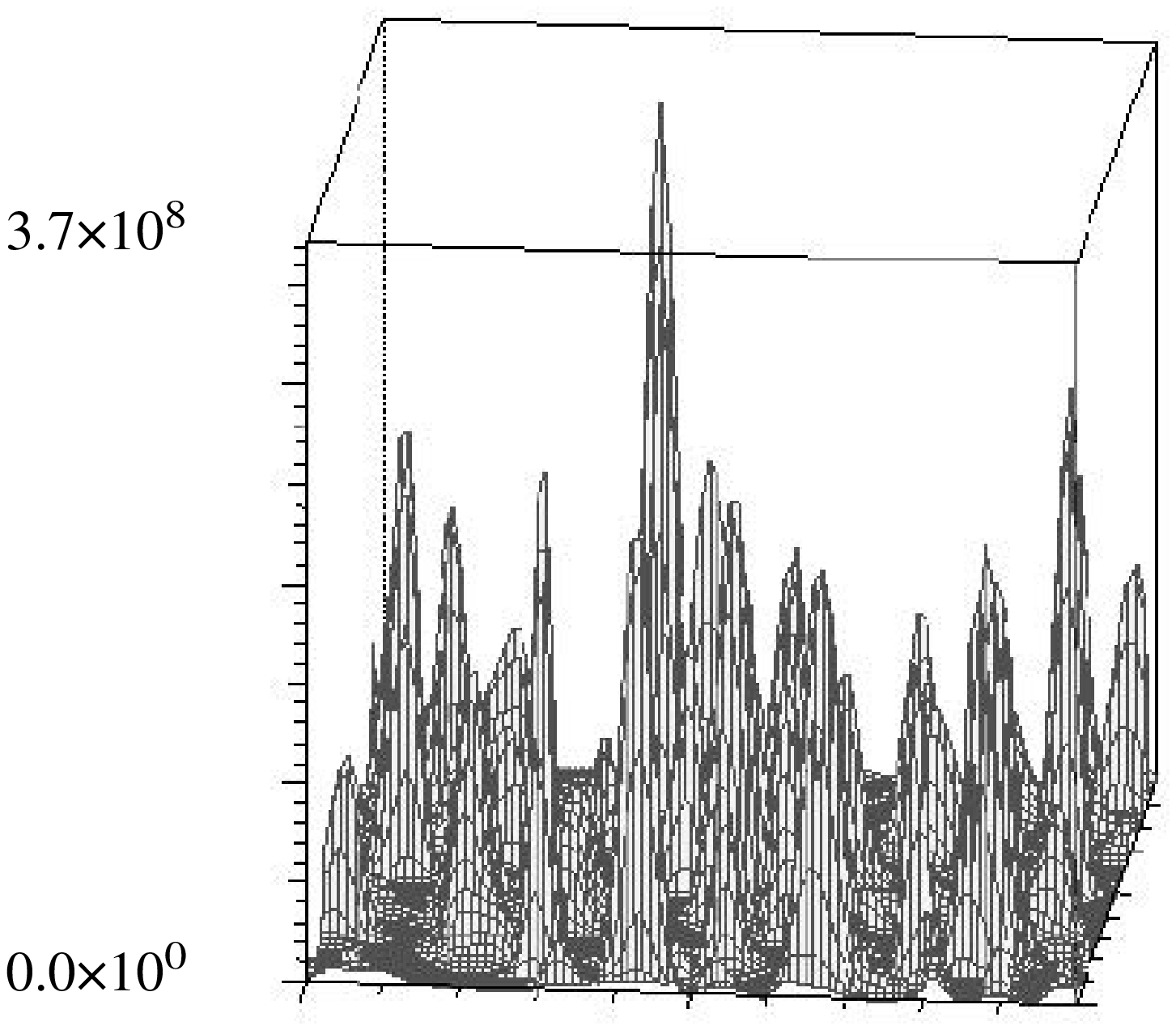}
\includegraphics{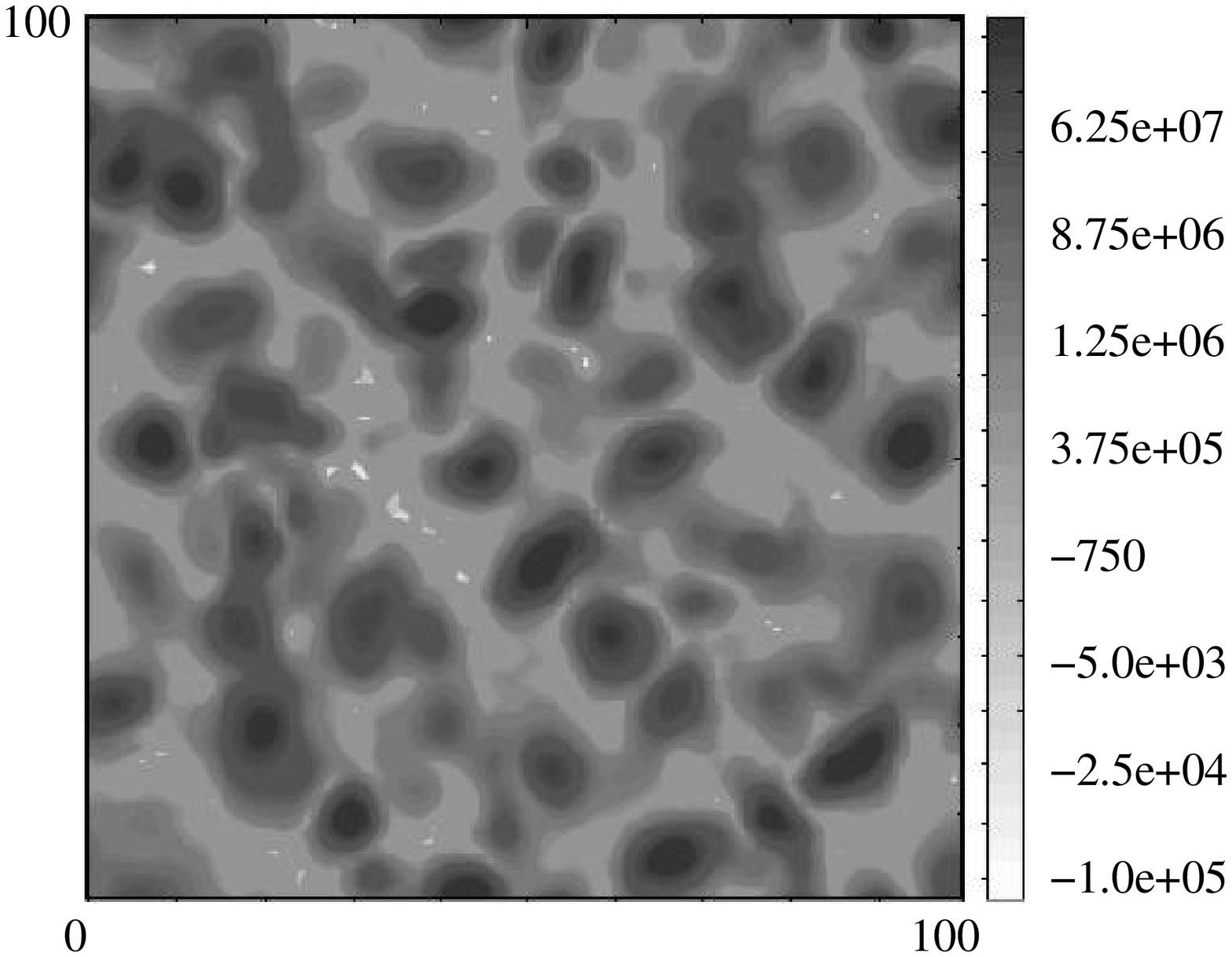}
\caption{$\tau=2610$} \label{form8} \vspace*{7cm}
\includegraphics{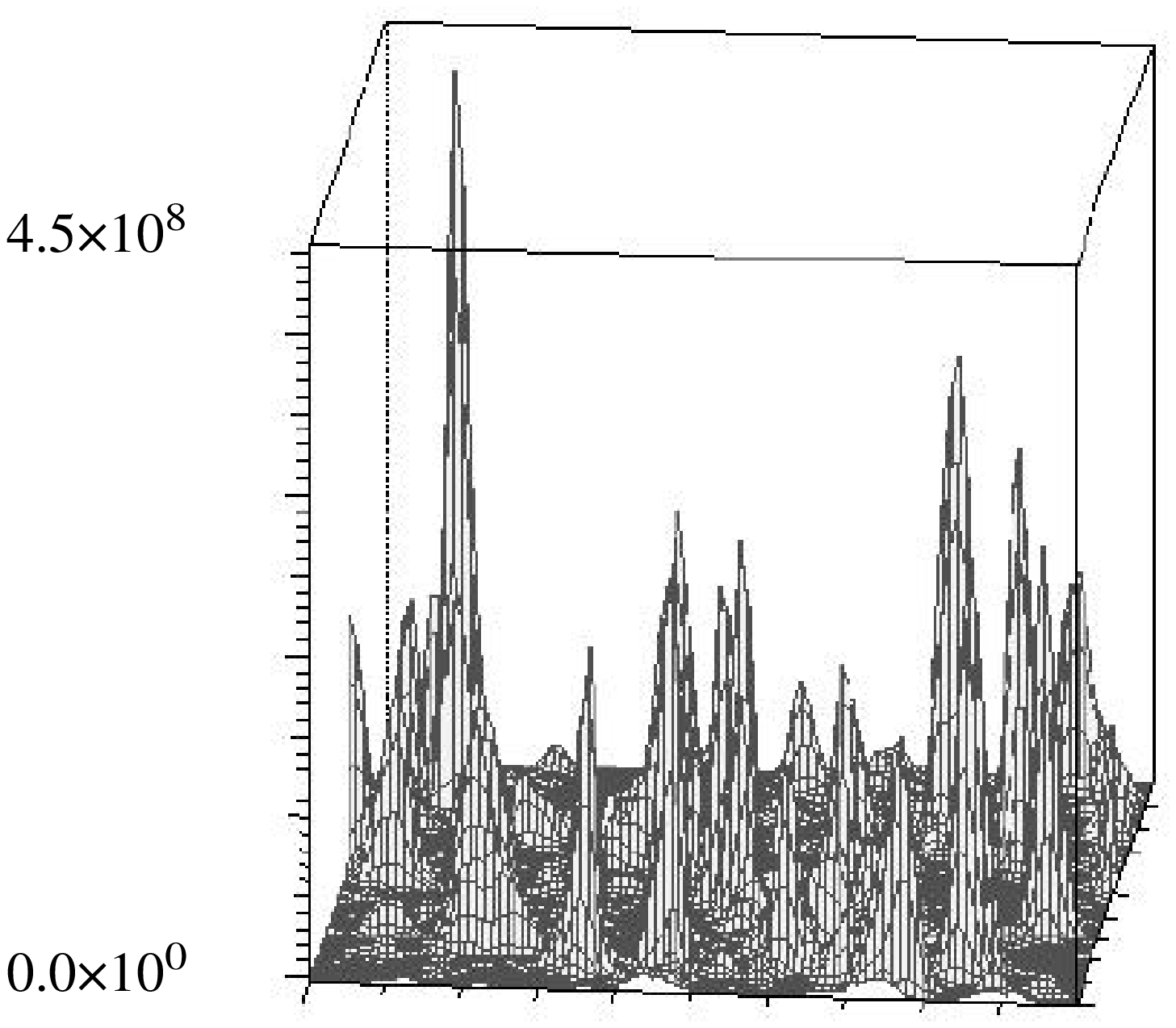}
\includegraphics{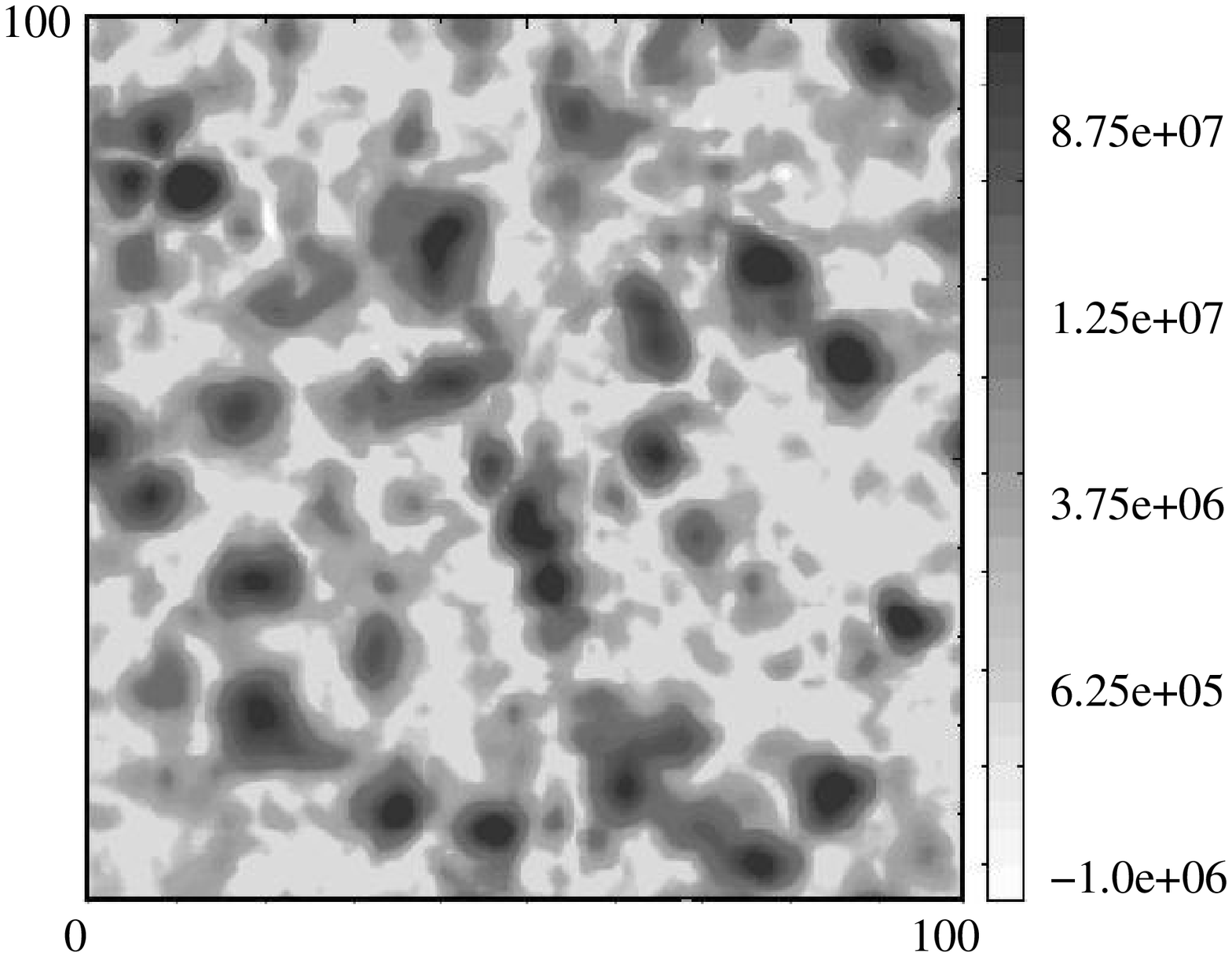}
\caption{$\tau=3060$}\label{form9}
\end{figure}

\begin{figure}[ht]
\leavevmode \centering \vspace*{7cm}
\includegraphics{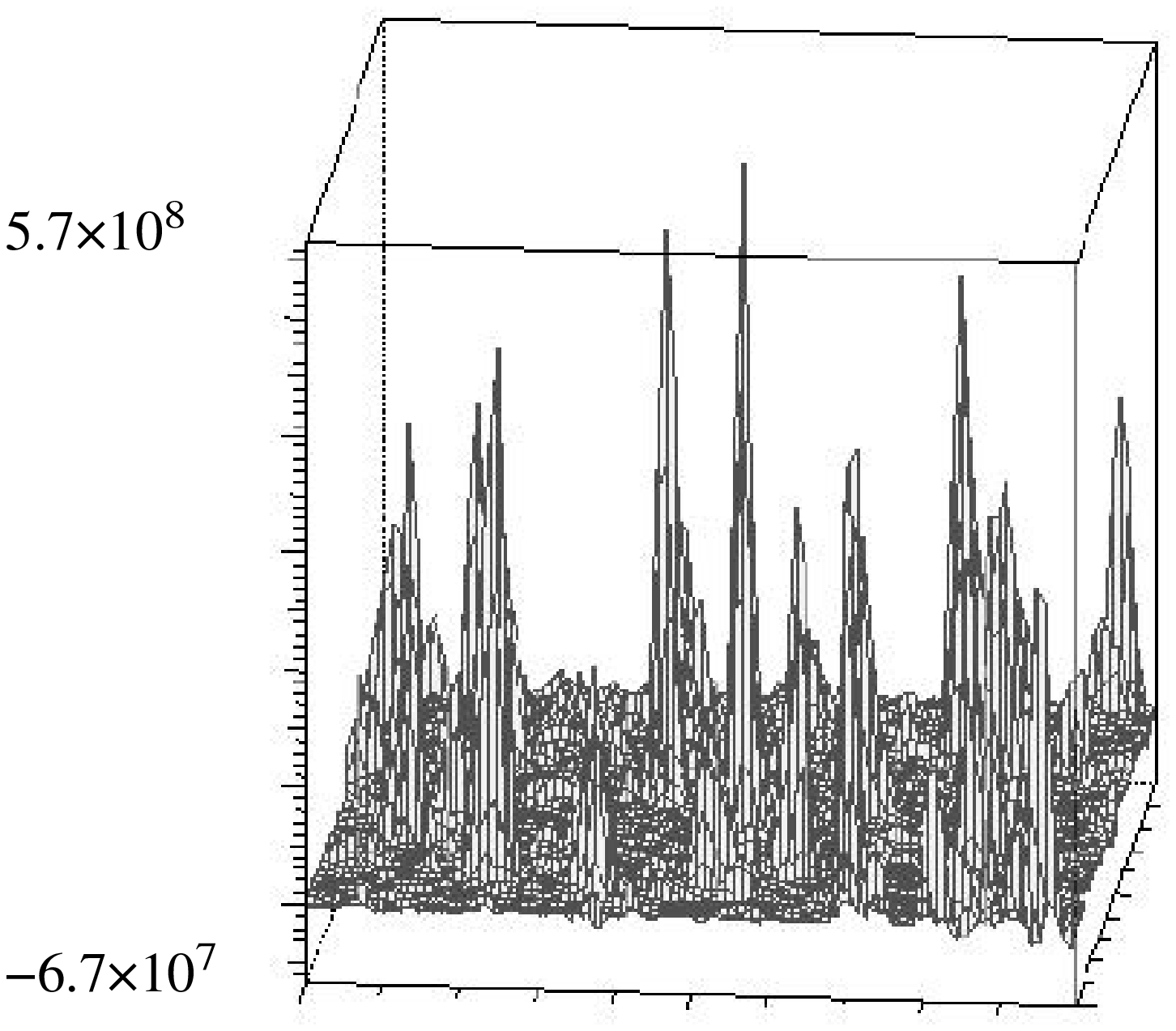}
\includegraphics{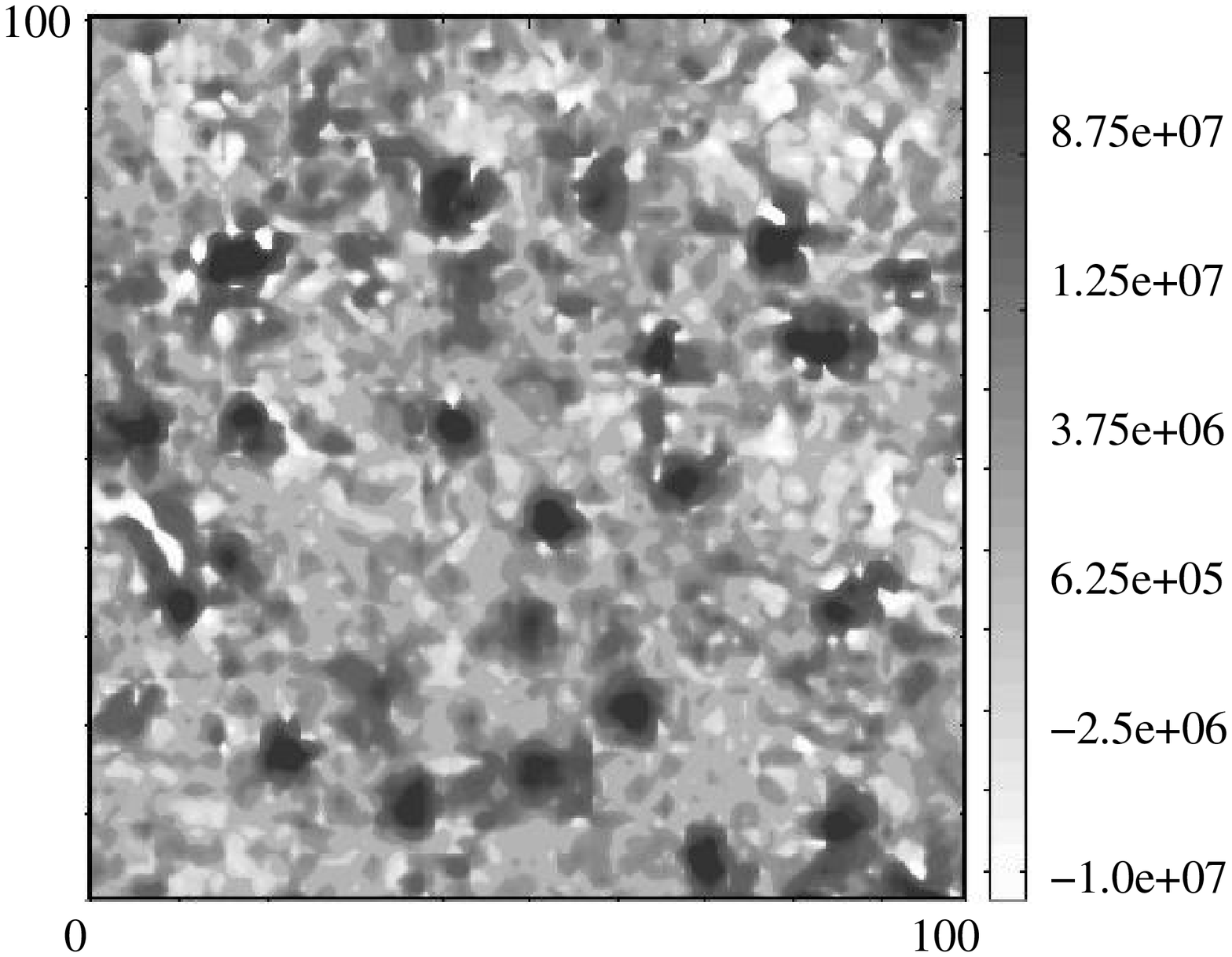}
\caption{$\tau=3600$}  \label{form10} \vspace*{7cm}
\includegraphics{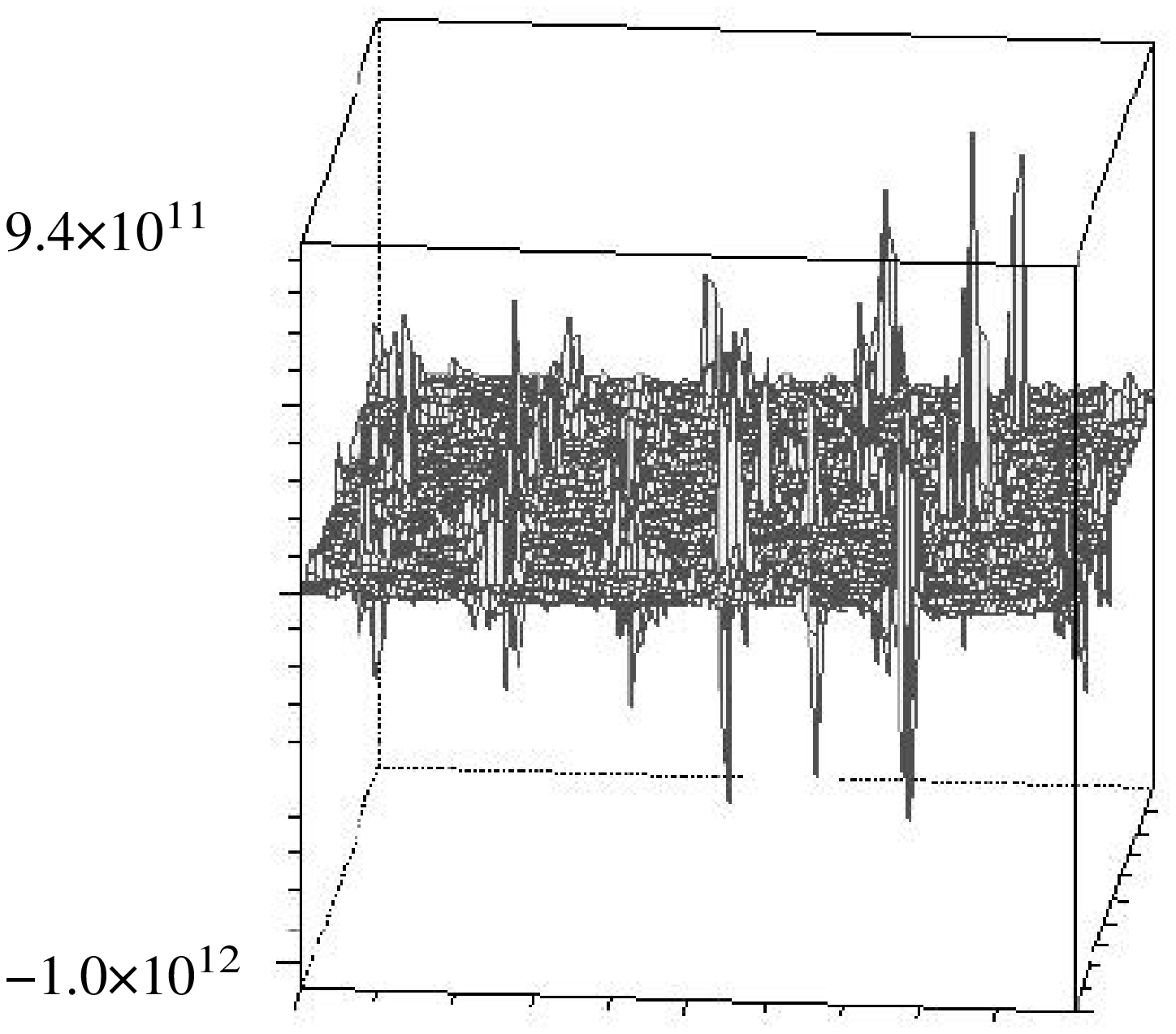}
\includegraphics{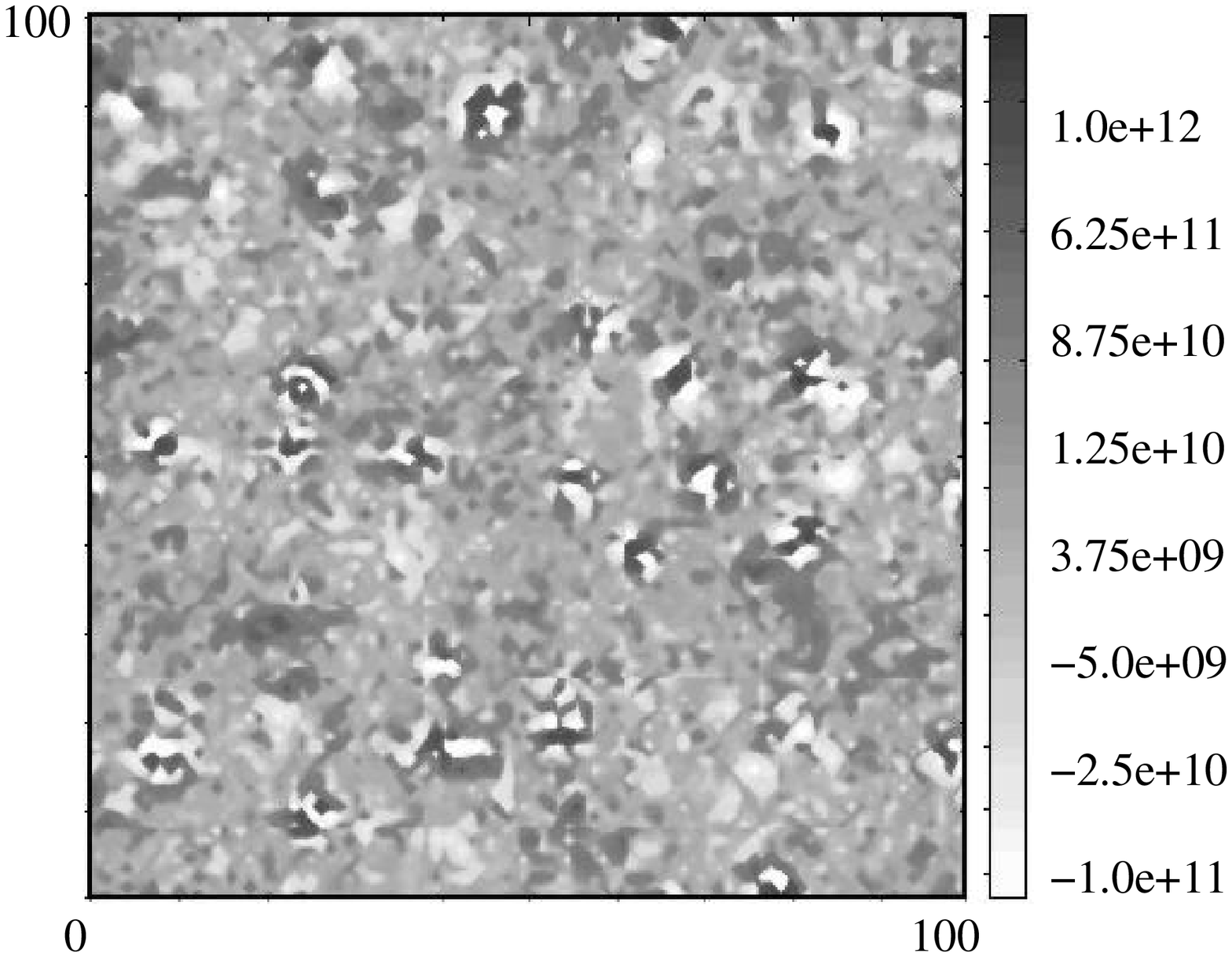}
\caption{$\tau=5400$}\label{form11}
\vspace*{7cm}
\includegraphics{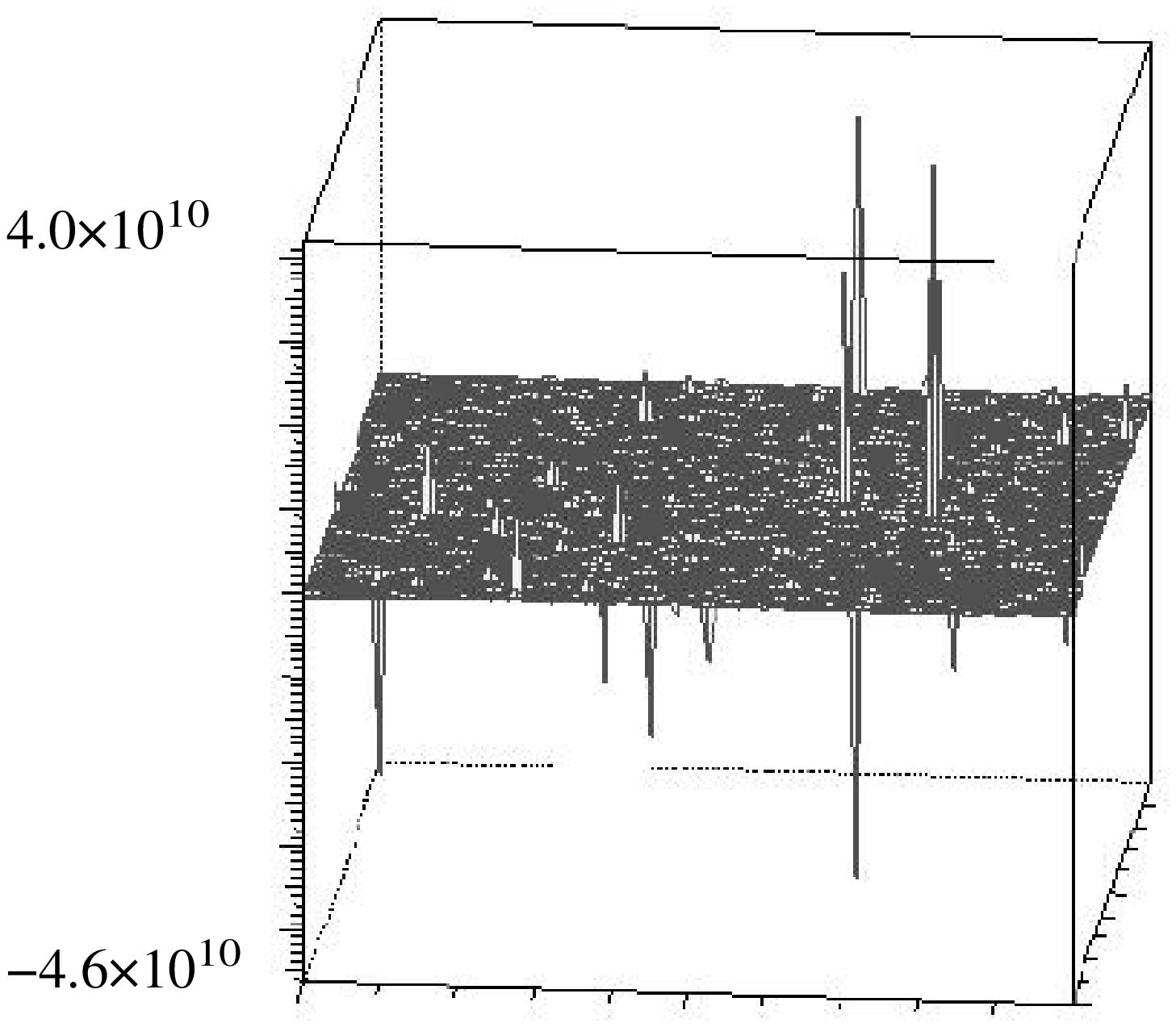}
\includegraphics{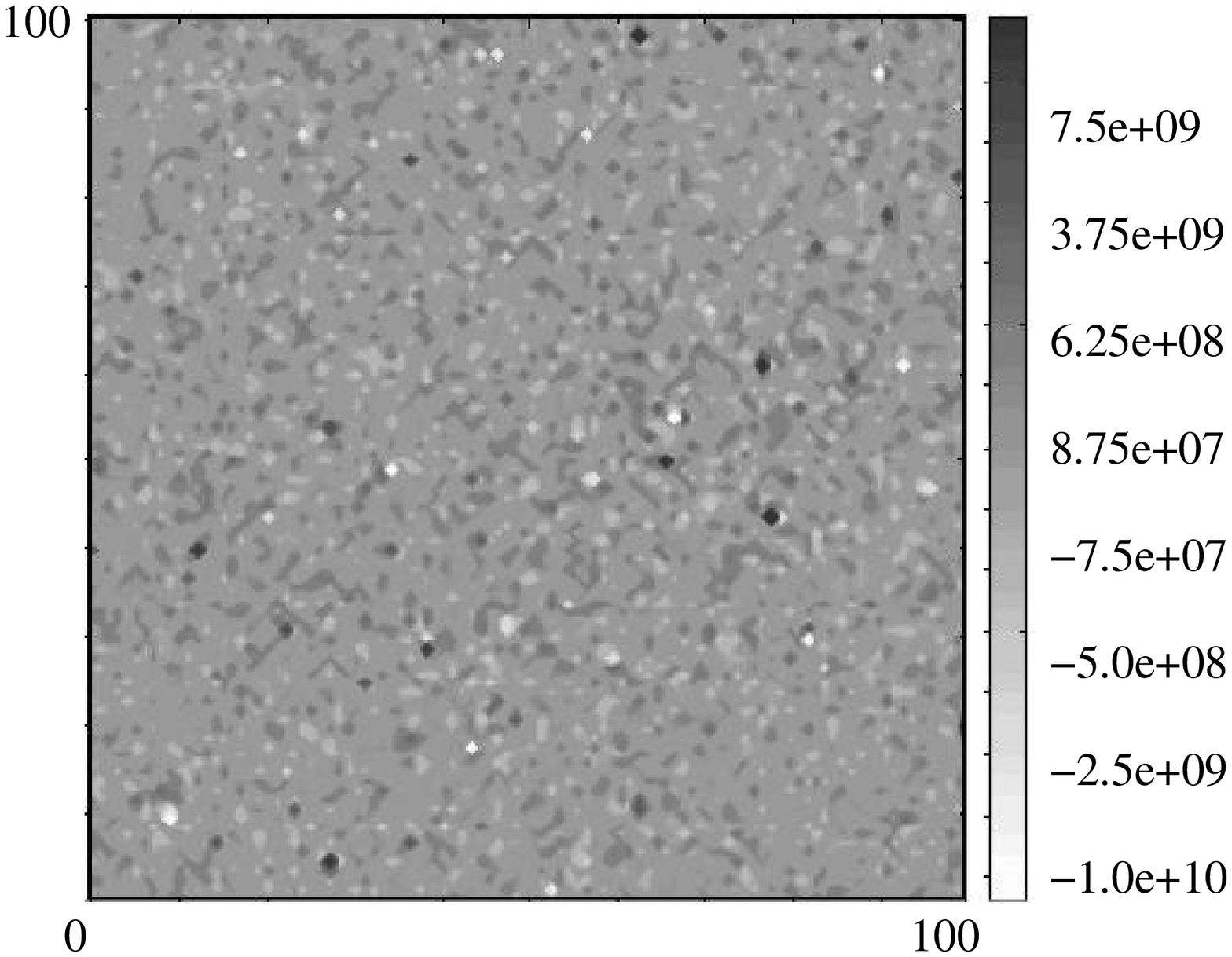}
\caption{$\tau=180000$}\label{form12}
\end{figure}

In Fig. \ref{form1} the initial small stochastic perturbations can
be seen clearly.
The fastest growing mode stars to dominate in Fig. \ref{form2}, while the small 
perturbations
are still visible. In Fig. \ref{form3} the growing modes are large enough  
for the initial perturbations to be no longer visible (they are obviously still present
but due to the increasing scale they simply can no longer 
be seen). The linear 
growth continues in Fig. \ref{form4} until non-linear growth begins
as depicted in Fig. \ref{form5}.
Rapid non-linear growth progresses and in Fig. \ref{form6} the fluctuations of the field
are already $\sim\cO(10^4)$ while up to Fig. \ref{form4} they 
are only at the level $\sim\cO(10^{-3})$.
In Fig. \ref{form6} we can see how the lumps of charge are 
dynamically arranged in string-like
features. These filaments are still visible in Fig. \ref{form7}, where
the condensate has further fragmented into lumps while the scale of the fluctuations 
has grown 
to $\sim\cO(10^{11})$. The filament texture
 disappears from sight in the next two figures, Figs \ref{form8} and
\ref{form9}, while negative charge starts to develop (some traces of the 
filaments can
still be seen in the distribution of charge lumps). 

Up to this point the cases $x=1$ and
$x\gg 1$ are qualitatively similar. If $x=1$, no negative charge develops and the
lumps slowly relax into Q-balls while the universe expands which finally freezes 
the distribution.
If, however, $x\gg 1$ like in the case shown here, new qualitative features become
apparent after this point. Negatively charged lumps start to become
visible in Fig. \ref{form10}, and in Fig. \ref{form11} the ratio of positive to 
negative charge
approaches one. The lumps of charge are complicated configurations that slowly relax
into Q-balls and anti-Q-balls. Finally, in Fig. \ref{form12} we can see separate
Q- and anti-Q-balls. The visible ten or so large Q-balls (anti-Q-balls) are only a 
very small
fraction of the total number of Q-balls present in the lattice. 

The different time scales of the fragmentation process can also be seen from
Figs \ref{form1}-\ref{form12}. The period of linear growth continues up to $\tau=2160$
after which rapid non-linear growth occurs between $\tau=2160 - 2520$. Negative
lumps begin to form at around $\tau=3600$ while the relaxation process to Q-balls
lasts until $\tau\sim 10^5$.

The process of Q-ball formation  has been illustrated in more detail
in Fig. \ref{formdetail}. We are again considering a 
comoving volume and the scale of the figures is varied between different panels.
The first panel is taken at $\tau=5400$ and the last panel at $\tau=180000$. 
Here $\w = 10^{-5}$
as before. In the first panel two large fluctuations that
both carry large positive and negative charge densities are
visible. Their shapes are still far from spherical and at this
stage the future evolution of the fluctuations is far from
obvious. In the following panels it can be
seen that the initial complex configurations more or less develop
into charge-anti charge -pairs that fluctuate and move with
time. Eventually only one large ball remains from each of the two
pairs. The final fate of the other partners  cannot be
determined from the Figures because of the presence of a large
background. It may be annihilated in the process, or it may move
away from the larger ball. It is clear, however, that the large
balls are formed from these complex disturbances through a
highly complicated relaxation process.

\begin{figure}[ht]
\leavevmode \centering \vspace*{65mm}
\includegraphics{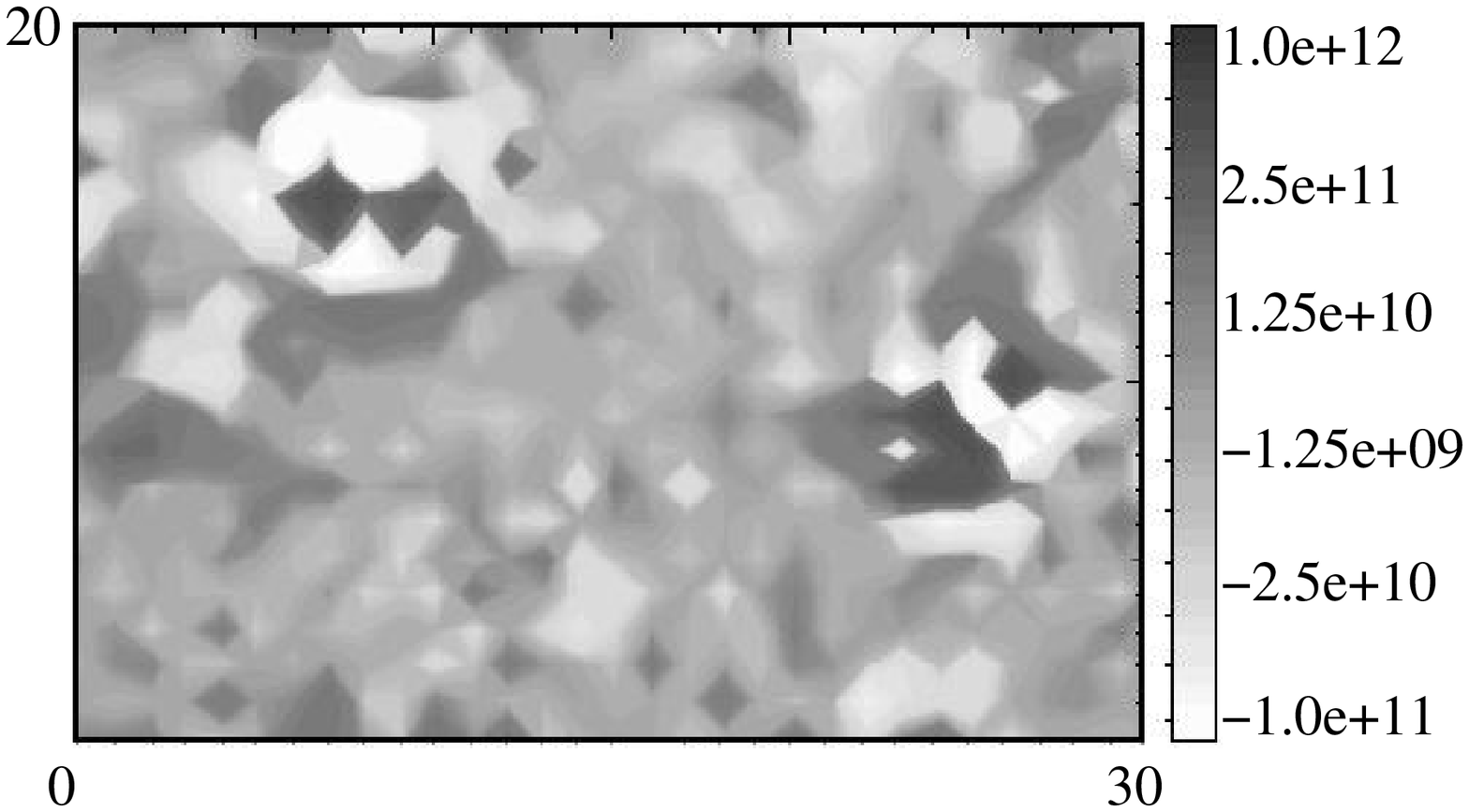}
\includegraphics{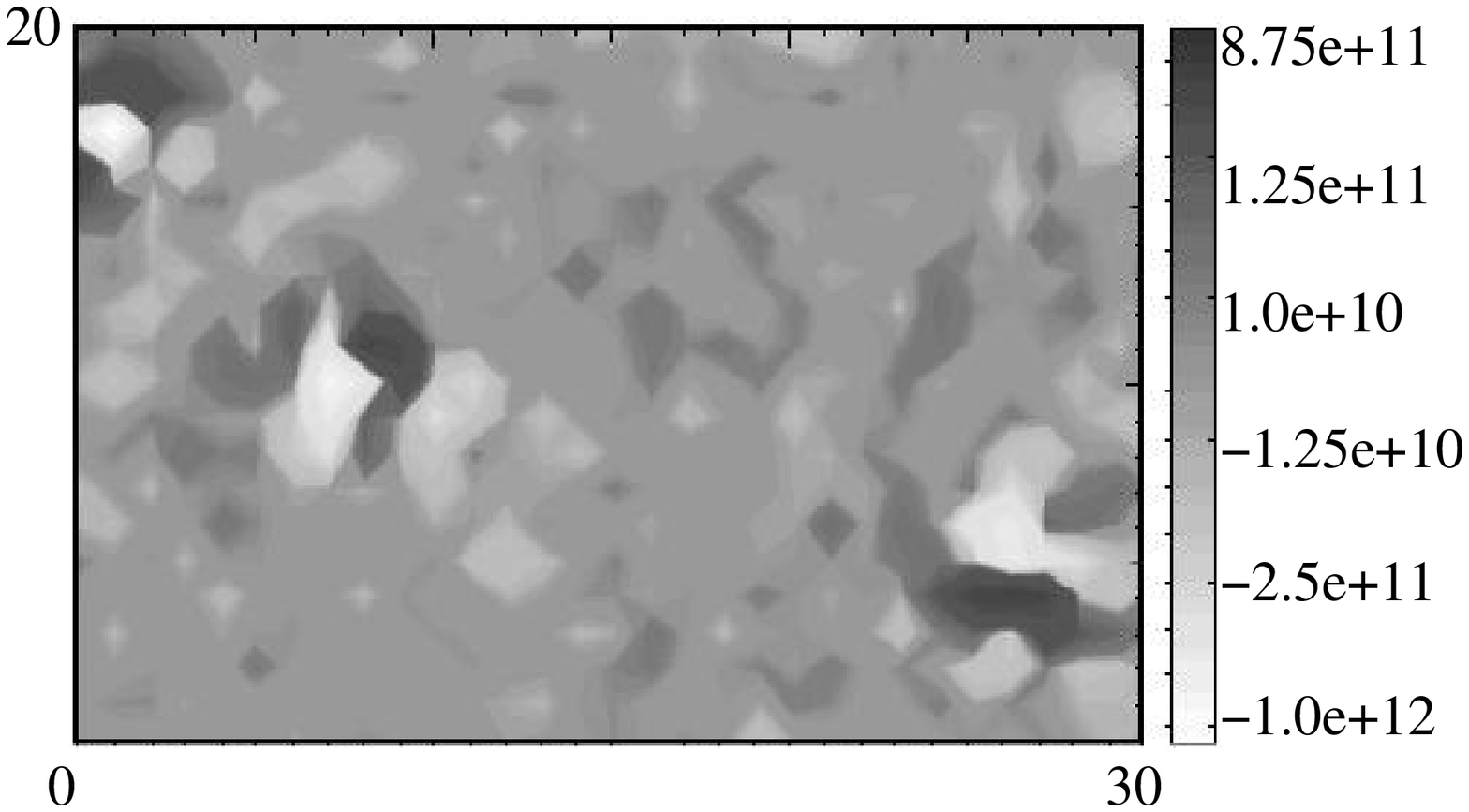}
\includegraphics{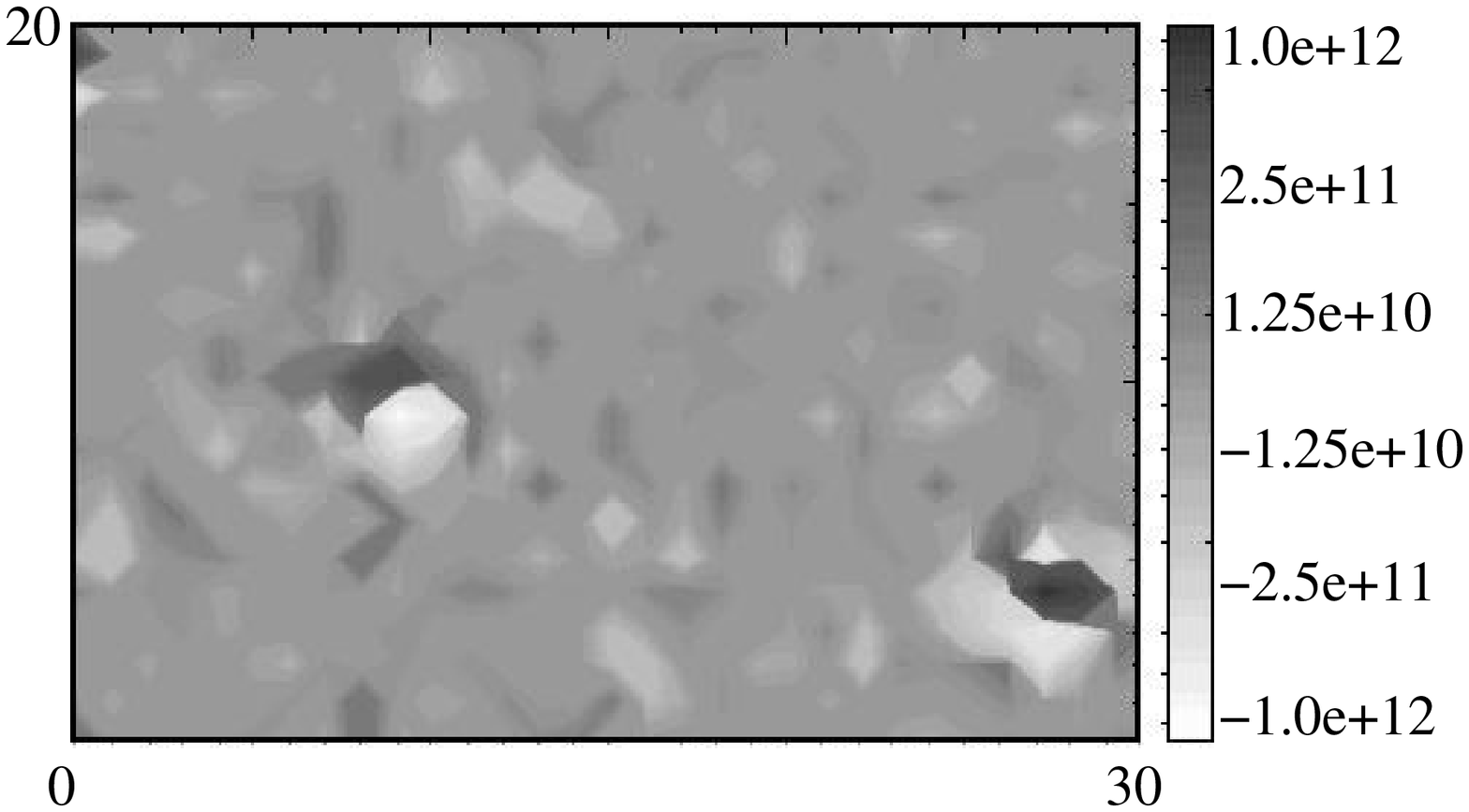}
\includegraphics{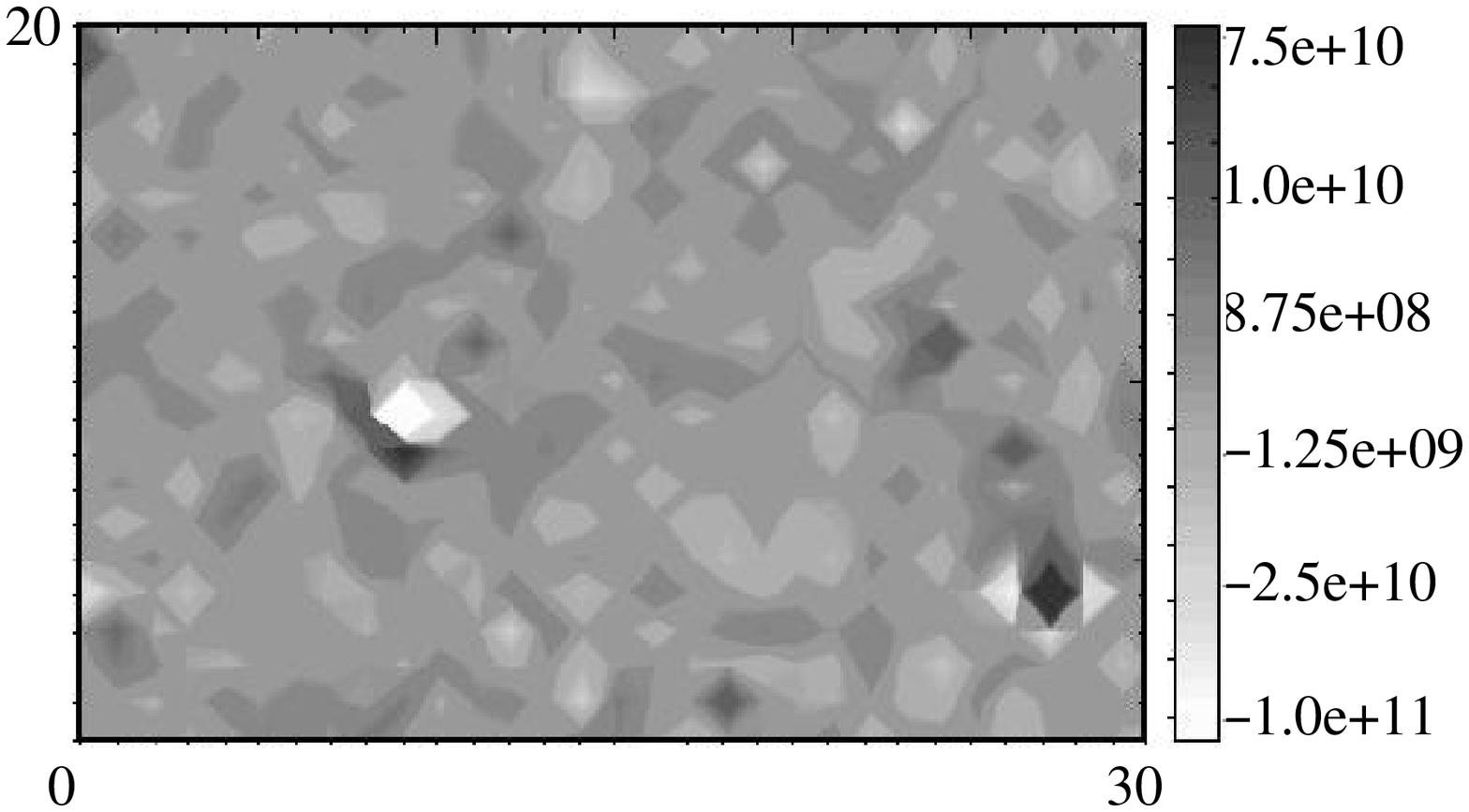}
\includegraphics{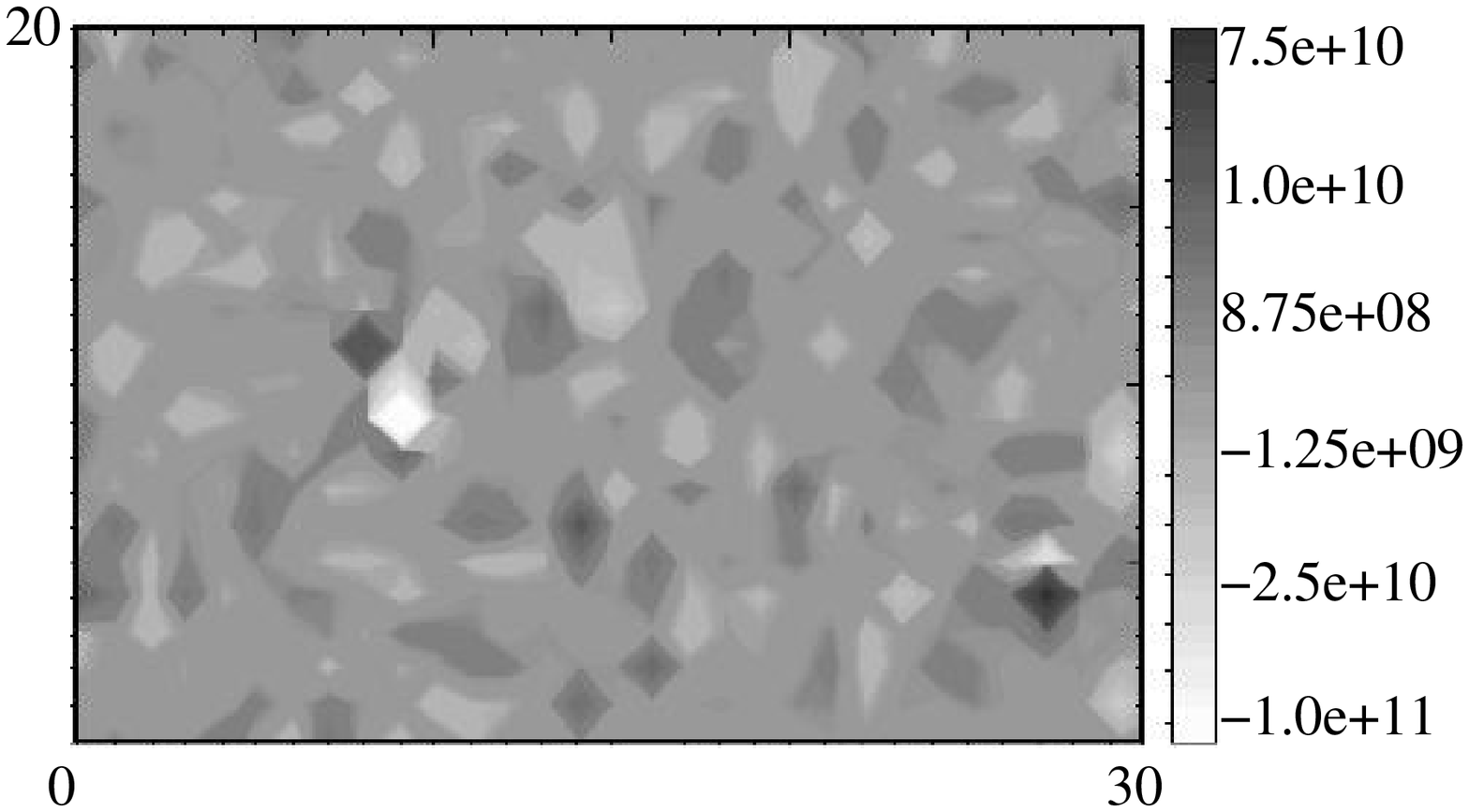}
\includegraphics{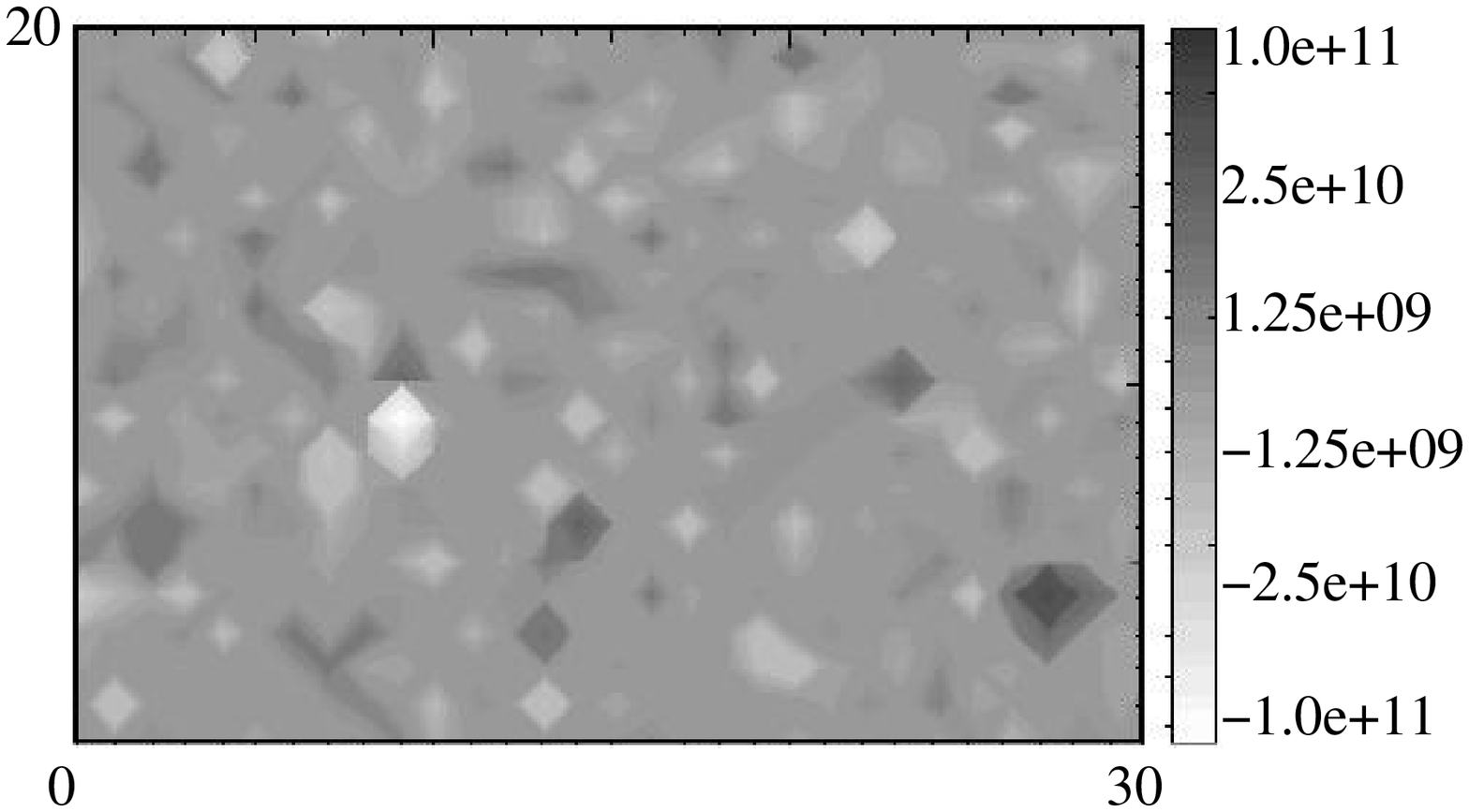}
\caption{A detail of the Q-ball formation process}\label{formdetail}
\end{figure}

\subsection{Distributions}

The simulations allow us to determine the charge distribution of
Q-balls and anti-Q-balls at different $\w$ and time intervals. In
practice we look for the distribution of the local charge maxima and
minima since the charge of a ball is approximately proportional
to these extremes. This is due to the fact that in the potential
studied in this work, the Q-ball radius has only a weak 
dependence on the charge and the profile can be well approximated 
with a gaussian {\it Ansatz} \cite{bbb2}.
The cumulative distributions are plotted in Fig. \ref{cdfs} for
the cases $\w=10^0,\ \w=10^{-5}$ with a detail of the $\w=10^{-5}$-case enlarged.

\begin{figure}[ht]
\leavevmode \centering \vspace*{12cm} 
\includegraphics{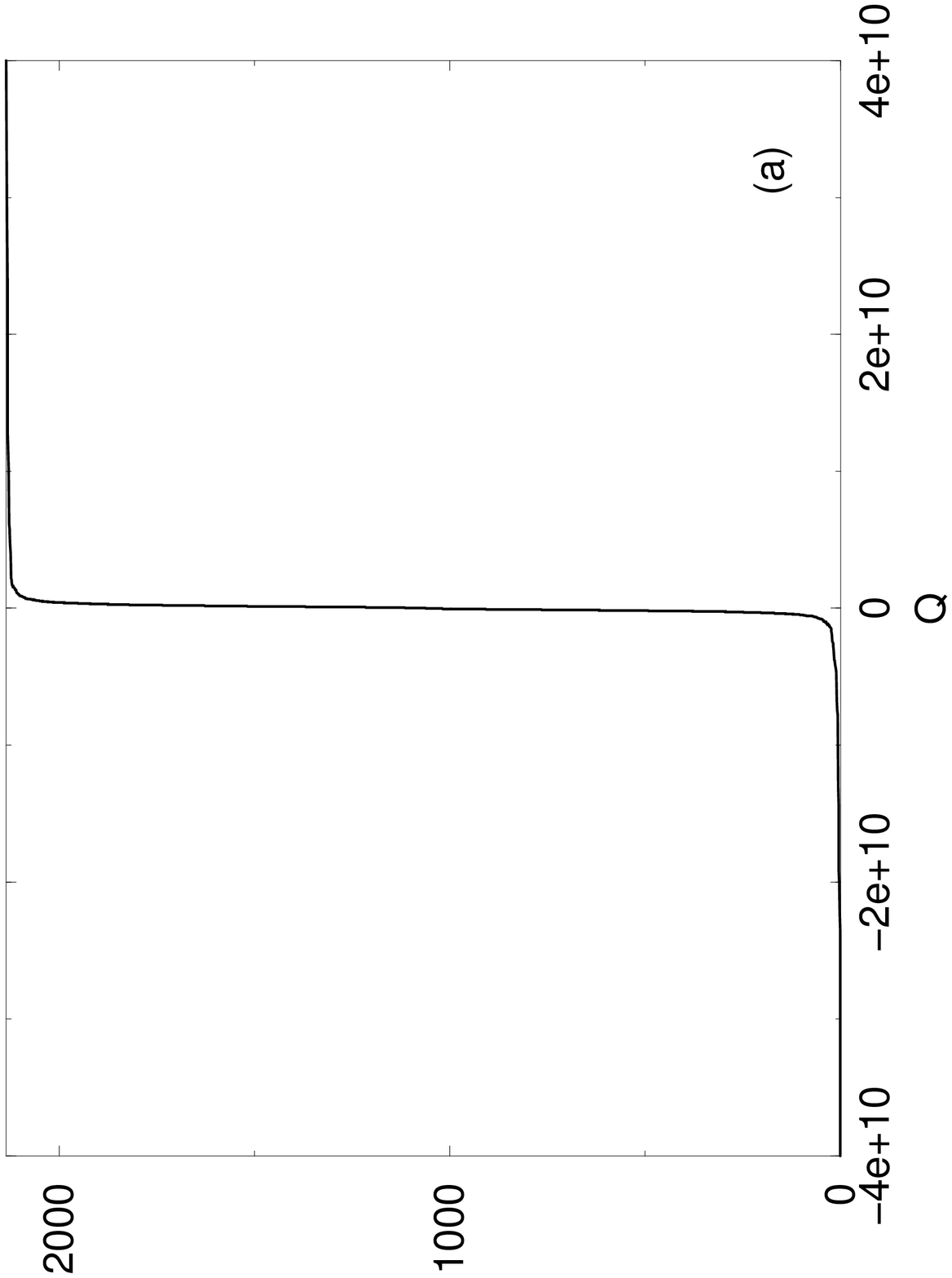} 
\includegraphics{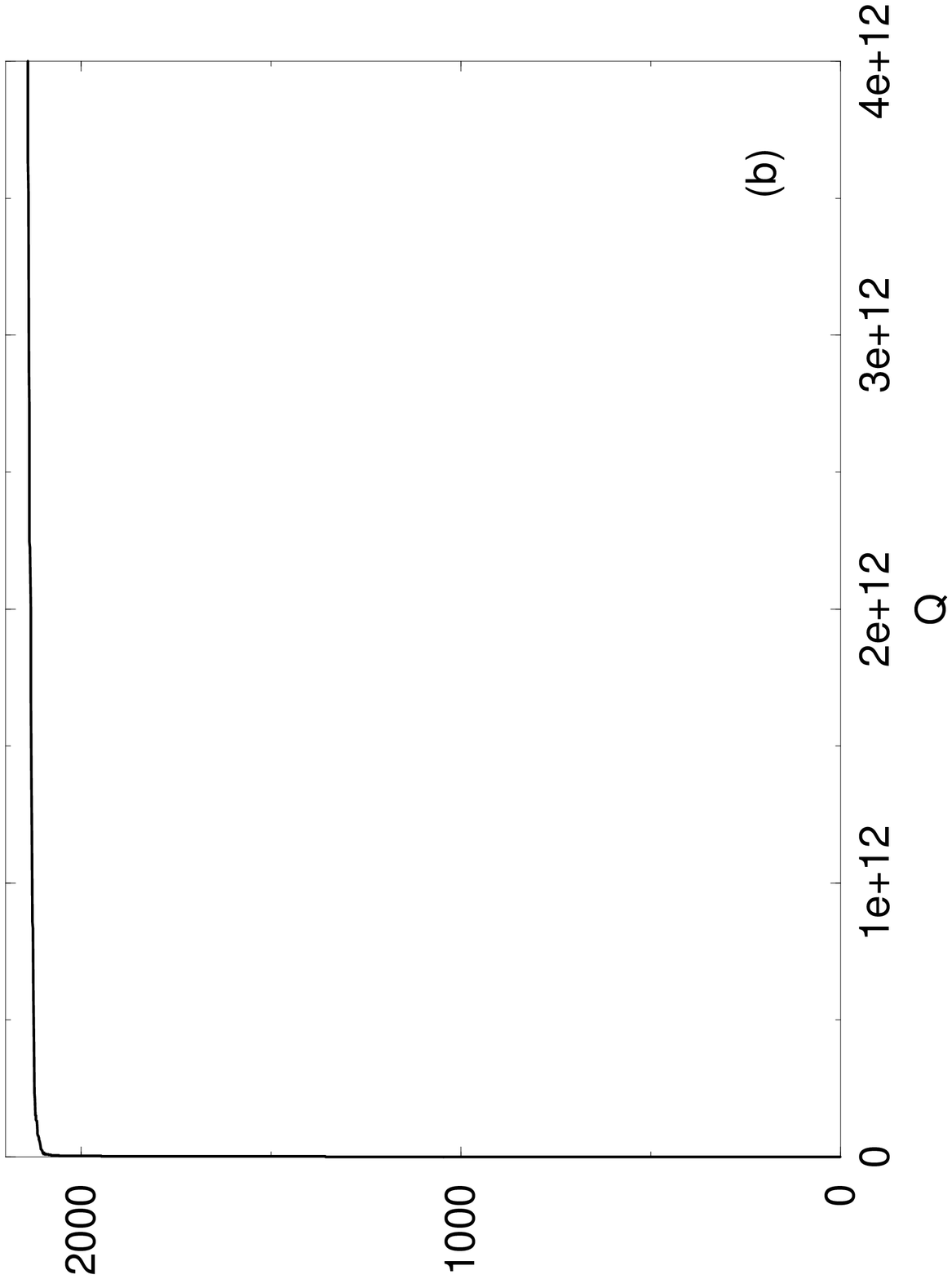} 
\includegraphics{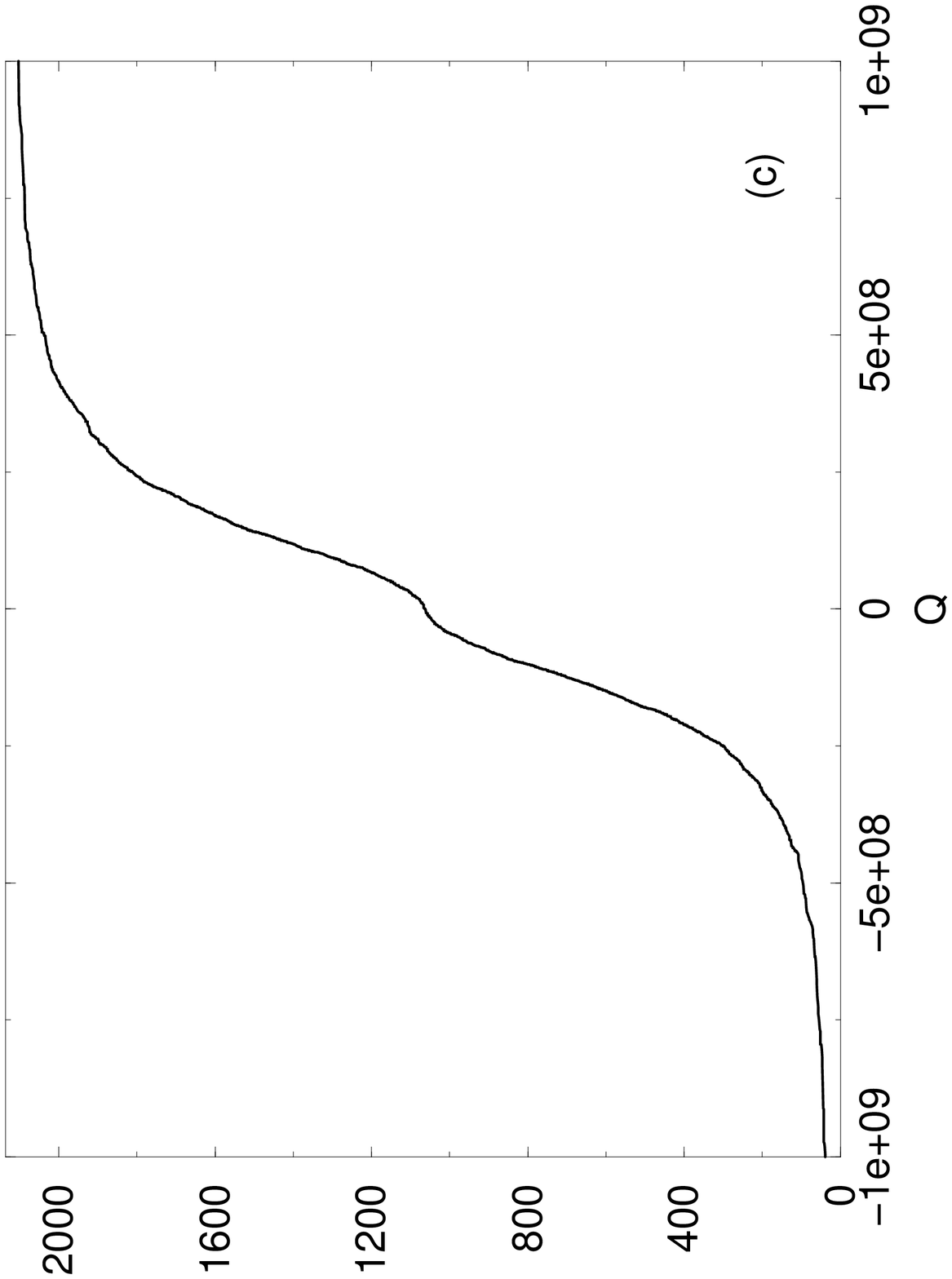}
\caption{Cumulative distributions for (a) $\w=10^{-5}$, (b) $\w=10^{0}$
and (c) a detail of the $\w=10^{-5}$-case}\label{cdfs}
\end{figure}

As can be seen from the Fig. \ref{cdfs}, the two cases are
quite different from each other as expected. In the $\w=10^0$ -case ($x\approx 1$) there form
practically only positive balls, whereas in the
$\w=10^{-5}$ -case an approximately equal number of positive and
negative balls are created. From the previous description of the
Q-ball formation process, these differing features are expected.
In the detail of the $\w=10^{-5}$ -case one can also note that there is
a small 'bump' at zero charge density.

We can now compare the distributions obtained from simulations to
the equilibrium distributions given in Sect. 2. The cumulative distribution
function, $F(Q,\mu,\beta)$, for a thermal distribution in
terms of the probability distribution function, $g(Q)=e^{\mu
Q-\beta |Q|}(\beta|Q|+1)$, reads as (omitting the fugacity of the 
distribution)
\be{CDF}
F(Q,\mu,\beta)=\integral{-\infty}{Q}dQ' g(Q')=\integral{-\infty}{Q}dQ' \e^{\mu
Q'-\beta |Q'|}(\beta|Q'|+1). \ee

However, as can be seen from Fig. \ref{cdfs}, the distribution
is not quite an equilibrium one but
deviates slightly from the thermal one because of
 the presence of the
'bump' at zero charge density. This feature is, however, a
lattice artifact, as we will see.

Let us account for the 'bump' by introducing a new
probability density function

\be{gnew} g(Q)=\left\{\begin{array}{ll}e^{\mu Q-\beta
|Q|}(\beta|Q|+1), & |Q|\geq Q_{\min}\\
0, & |Q|<Q_{\min},
\end{array} \right.
\ee where $Q_{\min}$ is a parameter to be determined by 
fitting to the numerical results. With this choice of the 
probability distribution function, the cumulative distribution 
function can be calculated to be

\be{fnew}
F(Q,\mu,\beta,Q_{\min})=y(\min\{Q,-Q_{\min}\})+y(Q_{\min})-
y(\max\{Q,Q_{\min}\}), 
\ee
where

 \be{ydef} y(Q)\equiv {Qe^{\mu Q-|Q|\beta}\over
Q\beta-|Q|\mu}(1+|Q|\beta+{Q\beta\over Q\beta-|Q|\mu}). 
\ee

We can now fit $y(Q)$ to the numerical data. An example is shown in 
Fig. \ref{fit} where $F(Q,\mu,\beta,Q_{\min})$ is fitted to the 
$\w=10^{-5}$-cases at $\tau=180000$. $y(Q)$ fits the data generally quite well
in all the cases where $x\gg 1$. If $\w=10^0$ one obviously needs to
consider a different type of fitting function since in this case no
anti-Q-balls are present in the distribution.

\begin{figure}
\leavevmode \centering \vspace*{10cm}
\includegraphics{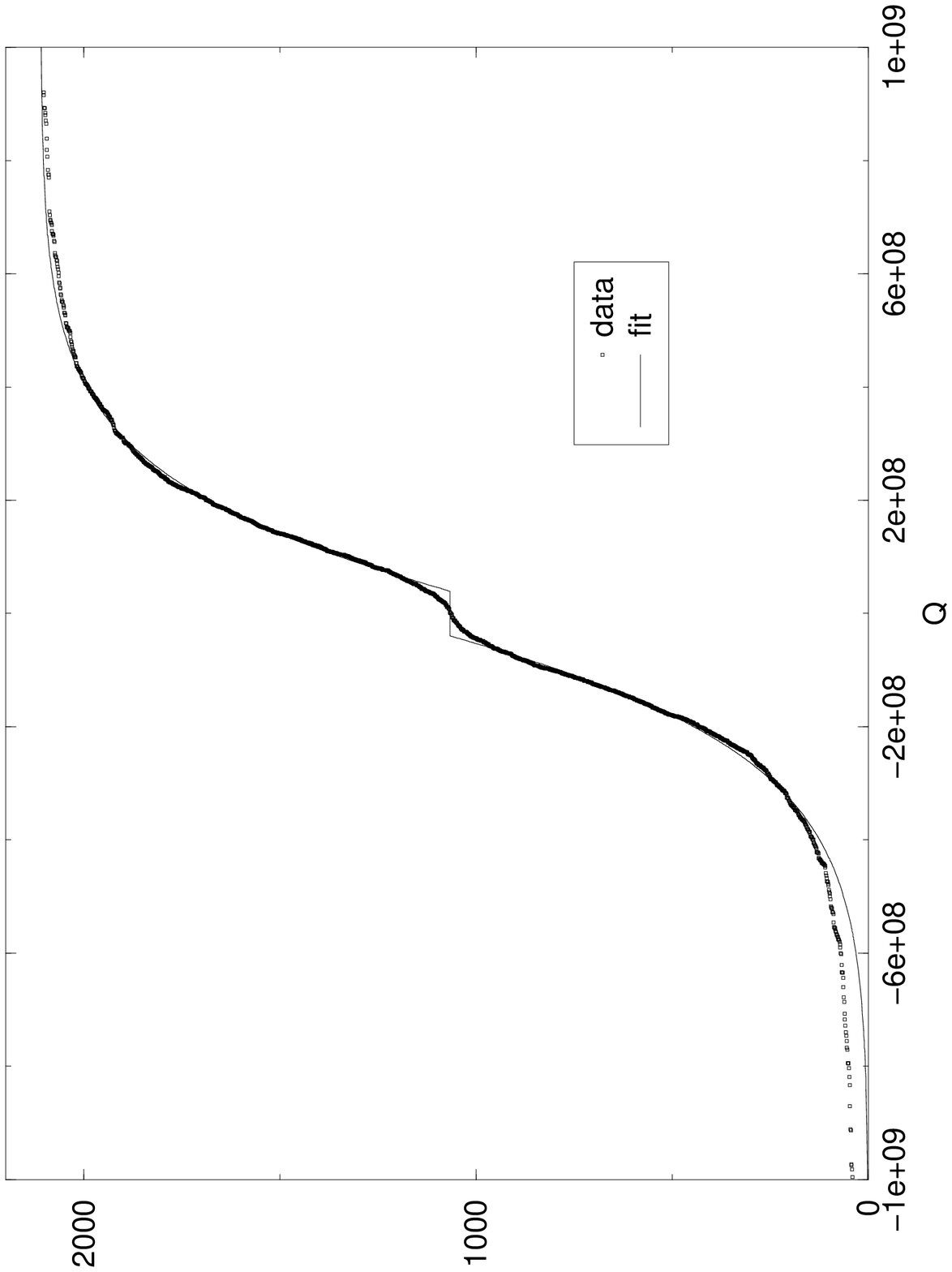}
\caption{A typical fit, $\w=10^{-5}$}\label{fit}
\end{figure}       

To test whether the inclusion of $Q_{\min}$ to the fitting
parameters is justifiable we have also fitted $F(Q,\mu,\beta)$ to
the data. In all the cases we found that this gives a much poorer
fit. In Fig. \ref{beta} we have plotted $({a_0/a})^2\beta$ at different 
time intervals for the $\w=10^{-5}$-case.

\begin{figure}[ht]
\leavevmode \centering \vspace*{10cm}
\includegraphics{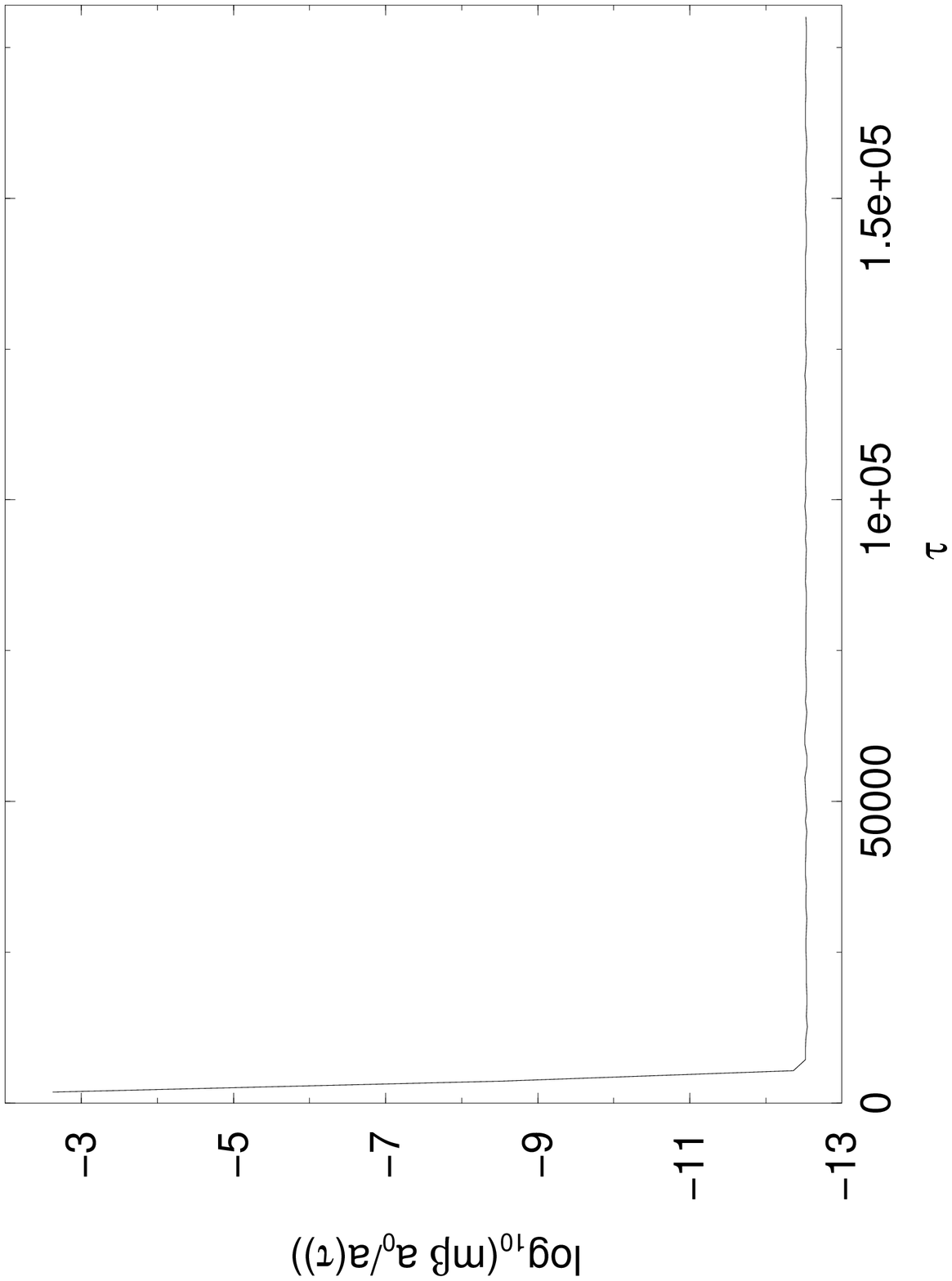}
\caption{$m\beta$ at different time intervals}\label{beta}
\end{figure}

The $({a_0/ a})^2$-factor is included to account for the expansion of the universe,
since in two dimensions the total energy density of Q-balls in equilibrium
behaves as $\sim 1/\beta$ for large $\beta $. On the other hand, the total
energy is essentially conserved so that $a(t)^2/\beta(t)$ remains a constant of
equilibrium.

{}From Fig. \ref{beta} it can be seen how $\beta$ quickly drops at around $\tau=3600-5400$.
This corresponds to Figures \ref{form10} and \ref{form11}. At this time the negative
Q-balls start to
form and the large initial charge lumps begin to fragment. The distribution
then quickly thermalizes and is well described by the equilibrium
distribution. We have run the simulation until $\tau=180000$ but in Fig.
\ref{beta} we have only plotted a part of the whole simulated time interval.
However, $({a_0/ a})^2\beta$ stays constant until the end of the
simulation. 

The presence of the 'bump' in the distribution may be attributed
to a number of factors. To study whether it is a lattice induced
effect we ran a number of simulations where the number of lattice
points was different while keeping the physical sized of the fiducial
volume
constant. After the fitting procedure we found that $Q_{\min}$
decreases with the number of lattice points used. Therefore it
is likely that the deviation from the simple thermal
spectrum is just due to lattice effects. This conclusion is 
further supported by the fact that in the gravity mediated case 
the charge-energy relation is approximately linear, $E\sim Q$, so 
that there is no preference for the formation of large Q-balls 
over small Q-balls\footnote{Note, that $Q_{min}$ is still a large number
$\sim 10^{7}$ so that the linear charge-energy -relation holds. For very small
$Q$ this relation is modified and therefore also the equilibrium
distribution for those values of $Q$ is not given by Eq. (\ref{CDF}). This does
not, however, affect our argument.}.  
In addition, when one considers the fact that the lattice spacing
increases with time and that smaller balls have smaller radii it is not too
surprising to see a small deviation from a purely thermal distribution
at small charge density. 

\section{Conclusions}

We have studied the process of Q-ball formation from the AD-condensate
in the early universe by numerical simulations in
two spatial dimensions. The analytical considerations 
presented in Sect. 2 reveal that 
a typical Q-ball is relativistic. This conclusion holds
if at least part of the baryon 
asymmetry of the universe is contained in baryon number carrying Q-balls.
 The reaction rate of Q-ball interactions 
is then much larger than the Hubble rate and hence we 
may expect that a state of maximum entropy
is likely to be reached. Assuming a thermal Q-ball distribution we 
showed that 
in a typical cosmological situation the AD-condensate fragments into 
practically equal numbers
of Q-balls and anti-Q-balls so that $N_+\approx N_-$. The  amount of charge 
contained in these balls is much larger than the net charge, 
$Q_+\approx Q_-\gg Q_{{tot}}$. Estimation of the average  
velocity of a Q-ball 
confirms that Q-balls are typically relativistic. The analytical calculations
were carried out in both two and three dimensions and we found
that the shapes of the number density distributions
or velocities are only very weakly dependent on dimension.
These considerations strongly support the assumption that 2-d
simulations can in fact capture the essential features of the
real 3-d case.

Our numerical simulations were carried out on a $2+1$ dimensional $100\times 100$-lattice. 
The initial condition was chosen to
be uniform AD-condensate with small
perturbations. The initial energy-to-charge ratio $x$ was varied from $1$ to $\sim 10^5$.
We found that regardless of $x$ the qualitative behavior
of the AD-condensate is similar in the beginning of the simulation. The small perturbations 
start to grow and due a complex dynamical process the AD-condensate 
fragments into several lumps carrying positive charge. 
These are not yet Q-balls but some excited states \cite{adfrag}.

{}From this point on the further evolution of the condensate depends on the actual value
of $x$. If $x=1$, the lumps slowly relax into Q-balls and no anti-Q-balls are present.
This is what we would
expect from analytical studies and this is what has been seen in 
3-d lattice simulations \cite{kasuya1,kasuya2}.
If, however, $x\gg 1$, the condensate fragments into a large
number of Q- and anti-Q-balls, as we have shown here. We also showed that 
the distributions of the Q- and anti-Q-balls obtained from the
simulations are thermal, which again is in accordance with 
analytical expectations.

An interesting feature which is apparent in Figs. \ref{form7}-\ref{form9}
is that the system of AD lumps
appears to possess some non-trivial spatial structure.
First, the large Q- and anti-Q-balls can be seen to form in pairs which 
give rise to a positive spatial
correlation between them. It should be noted that if small balls are also
formed in pairs they are likely to have a homogenized spatial distribution due to their
large velocities. Second,  there are long range correlations
which show up as filament-like structures in Figs. \ref{form6}-\ref{form8}.
The texture visible in the middle phase of the non-linear evolution
may well survive in the spatial Q-ball distribution also at later
times, although it no longer is visible. Numerical correlation analysis indeed
shows that still at the final stage of our simulation (Fig. \ref{form9})
the mean distance between both Q- and anti-Q-balls deviates by over four
standard deviations from the random uniform distribution. The deviation
is negative \ie the mean distance between Q-balls is smaller than expected
for a random distribution. These features may affect
observables such as the cosmic microwave background or large scale structure
and deserve further study.

The fact that the Q-ball distribution in the cosmologically realistic case, $x\gg 1$,
tends toward a thermal one implies, like the analytical treatment shows, that a large
number of Q-balls move relativistically. The very large Q-balls that are clearly visible
in Fig. \ref{form12} appear immobile but there is also 
a large number of smaller fast
moving Q-balls which can thermalize the distribution. In a
thermodynamical sense one can view the
fragmentation of the AD-condensate into Q-balls as a relaxation process where entropy
is maximized: from this point of view the appearance of the thermal distribution is natural.

In this work we left the gauge mediated SUSY breaking scenario without the attention it
deserves. This is due to a number of facts that make 
the simulation of the gauge mediated
SUSY breaking scenario more cumbersome compared to the gravity mediated case. 
In the fragmentation process
the spectra of growing modes play an
important role since the first mode that goes non-linear determines the size of the fastest
growing disturbance in the condensate. The SUSY breaking mechanism, \ie whether it is gauge or
gravity mediated, determines the size of the fastest growing mode via the actual form
of the scalar potential. As we
have seen, the size of the instabilities in the gauge mediated case are larger than in the
gravity mediated scenario. Therefore one needs a lattice with a much larger physical
size. Obviously,
one can always increase lattice spacing but 
this happens at the cost of losing resolution. Moreover,
the large instabilities correspond to large Q-balls forming from the 
AD-condensate which in
the gauge mediated scenario have thin walls. To avoid numerical
instabilities
due to the thin walls a high resolution is required.
Therefore, in order to 
avoid finite size effects, one would need a much larger number of lattice points than
in the gravity mediated case.

In addition to the numerical difficulties the gauge mediated scenario does
not benefit form the analytical
arguments presented in Sect. 2.
This is due to the different
energy-charge -relations of the two cases. For large Q-balls in the gauge mediated case the
energy is related to the charge by $E\sim Q^{3\over 4}$ instead of the
linear relation $E\sim Q$ in the gravity mediated scenario. From (\ref{part1}) one can see that the
Q-integral diverges since the $\exp (-\beta E+\mu Q)$-term is dominated by $\mu Q$ in the
large Q limit in the gauge mediated scenario. Physically the lack of a
proper asymptotic
equilibrium state implies that the system tends
towards a state where charge is concentrated in large balls (and ultimately
in a single Q-ball) as noted also in
Ref.\ \cite{ls}. However, one should also
take into account that the more charge is stored in large
balls the more there exists leftover energy in the system that must be
accounted for. The final state of the system would then be strongly dependent
on the time at which the distribution freezes due to the expansion of the
universe and hence on the initial conditions.

 To conclude, our results
indicate that in the gravity mediated SUSY breaking case, the AD condensate
fragments into an almost equal number of  Q-balls and anti-Q-balls.
The evaporation times of the Q-balls, the nature of particles they decay into,
and possible spatial correlations surviving the fragmentation process
and the subsequent decays of Q-balls, are
obviously very important factors that determine the signatures left by 
a possible Q-ball
period in the early history of the universe. 


\section*{Acknowledgements}

This work is partly supported by the Academy of Finland under the
contract 101-35224 and by the Finnish Graduate School of Particle
and Nuclear Physics.

\end{document}